\documentclass[letter]{aa} % for the letters 
\usepackage{graphicx}
\usepackage{natbib}
\usepackage{longtable,lscape}
\usepackage{txfonts}
\begin{document}
\title{FIR-detected Lyman break galaxies at $z \sim 3$\thanks{{\it Herschel} is an ESA space observatory with science instruments provided by European-led Principal Investigator consortia and with important participation from NASA.}}
\subtitle{Dust attenuation and dust correction factors at high redshift}

\authorrunning{Oteo et al. 2013}
\titlerunning{FIR-detected LBGs at $z \sim 3$}

\author{
I. Oteo \inst{1,2},
J. Cepa \inst{1,2},
\'A. Bongiovanni \inst{1,2,3},
A. M. P\'erez-Garc\'ia \inst{1,2,3},
B. Cedr\'es \inst{1,2},
H. Dom\'inguez S\'anchez \inst{1,2},
A. Ederoclite \inst{4}, 
M. S\'anchez-Portal \inst{3,5}, 
I. Pintos-Castro \inst{1,2,6}, and 
R. P\'erez-Mart\'inez \inst{7}\\
}
\offprints{Iv\'an Oteo, \email{ioteo@iac.es}}

\institute{Instituto de Astrof{\'i}sica de Canarias (IAC), E-38200 La Laguna, Tenerife, Spain
\and Departamento de Astrof{\'i}sica, Universidad de La Laguna (ULL), E-38205 La Laguna, Tenerife, Spain
\and Asociaci\' on ASPID. Apartado de Correos 412, La Laguna, Tenerife, Spain.
\and Centro de Estudios de F\'isica del Cosmos de Arag\' on, Plaza San Juan 1, Planta 2, Teruel, 44001, Spain
\and Herschel Science Centre (ESAC). Villafranca del Castillo, Spain
\and Centro de Astrobiolog\'{i}a, INTA-CSIC, P.O. Box - Apdo. de correos 78, Villanueva de la Ca\~nada Madrid 28691, Spain
\and XMM/Newton Science Operations Centre (ESAC). Villafranca del Castillo. Spain
}

\date{Received ...; accepted... }

% \abstract{}{}{}{}{} 
% 5 {} token are mandatory

 \abstract
   {Lyman break galaxies (LBGs) represent one of the kinds of star-forming galaxies that are found in the high-redshift universe. The detection of LBGs in the FIR domain can provide very important clues on their dust attenuation and total SFR, allowing a more detailed study than those performed so far. In this work we explore the FIR emission of a sample of 16 LBGs at $z \sim 3$ in the GOODS-North and GOODS-South fields that are individually detected in PACS-100$\mu$m or PACS-160$\mu$m. These detections demonstrate the possibility of measuring the dust emission of LBGs at high redshift. We find that PACS-detected LBGs at $z \sim 3$ are highly obscured galaxies which belong to the Ultra luminous IR galaxies or Hyper luminous IR galaxies class. Their total SFR cannot be recovered with the dust attenuation factors obtained from their UV continuum slope or their SED-derived dust attenuation employing \cite{Bruzual2003} templates. Both methods underestimate the results for most of the galaxies. Comparing with a sample of PACS-detected LBGs at $z \sim 1$ we find evidences that the FIR emission of LBGs might have changed with redshift in the sense that the dustiest LBGs found at $z \sim 3$ have more prominent FIR emission, are dustier for a given UV slope, and have higher SFR for a given stellar mass than the dustiest LBGs found at $z \sim 1$.
   }

   \keywords{cosmology: observations --
                galaxies: stellar populations, morphology, infrared.
               }

   \maketitle

%________________________________________________________________

\section{Introduction}\label{intro}

The dropout technique, that segregates the so-called Lyman break galaxies (LBGs), is one of the most employed and successful methods to find high-redshift star-forming (SF) galaxies \citep{Burgarella2006,Burgarella2007,Chen2013,Oteo2013a,Basu2011,Hathi2012,Madau1996,Steidel1996,Steidel1999,Steidel2003,Stanway2003,Giavalisco2004LBGs,Bunker2004,Verma2007,Iwata2007}. Many studies have analyzed the physical properties of LBGs with photometric and spectroscopic data, from X-ray to radio wavelengths. The detection rate of LBGs is very dependent upon their redshift and the wavelength of the observations. LBGs can be easily detected in the optical with the current deep cosmological surveys. A fraction of them are detected near-IR emission. The situation becomes problematic when we move to redder wavelengths and study high-redshift ($z \geq 3$) LBGs. Only massive LBGs at $z \sim 3$, $\log{\left( M/M_\odot \right)} \sim 11$, are detected in IRAC or MIPS channels \citep{Magdis2010}. The detection of some LBGs at different redshifts with MIPS-24$\mu$m revealed a population of luminous infrared LBGs \citep{Huang2005,Rigopoulou2006,Burgarella2006,Burgarella2007,Basu2011}. On the FIR side, \cite{Burgarella2011} found SPIRE detections for a sample of 12 GALEX-selected LBGs at $z \sim 1$ and one LBG at $z \sim 2$. \cite{Rigopoulou2010} did not found individually detected LBGs at $z \sim 3$ in SPIRE bands and \cite{Magdis2010LBGs} did not found PACS-FIR counterparts of their LBGs at $z \sim 3$ either. In redder wavelengths than FIR, very few LBGs have been directly detected \citep{Chapman2000,Chapman2009}. Recently, \cite{Magdis2012_CO} present CO[3$\rightarrow$2] observations of two PACS-detected FIR-bright LBGs at $z \sim 3$ which indicate that the steep evolution of $M_{\rm gas}/M_*$ of normal galaxies up to $z \sim 2$ is followed by a flattening at higher redshifts, providing evidence for the existence of a plateau in the evolution of the specific star formation rate (sSFR) at z > 2.5 \citep[see also][]{2012ApJ...754...83B}.

\begin{figure*}
\centering
\includegraphics[width=0.2\textwidth]{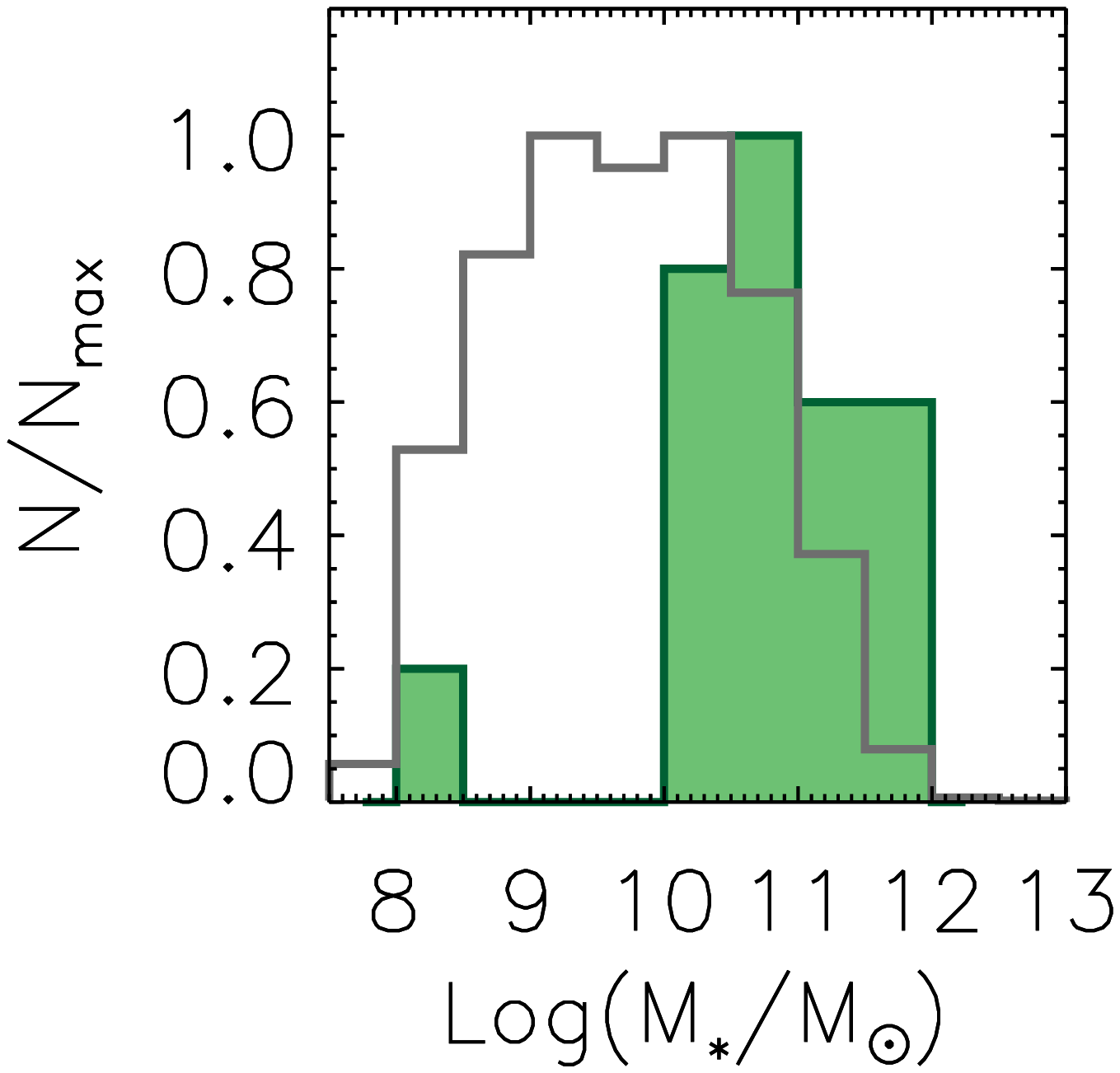}
\includegraphics[width=0.2\textwidth]{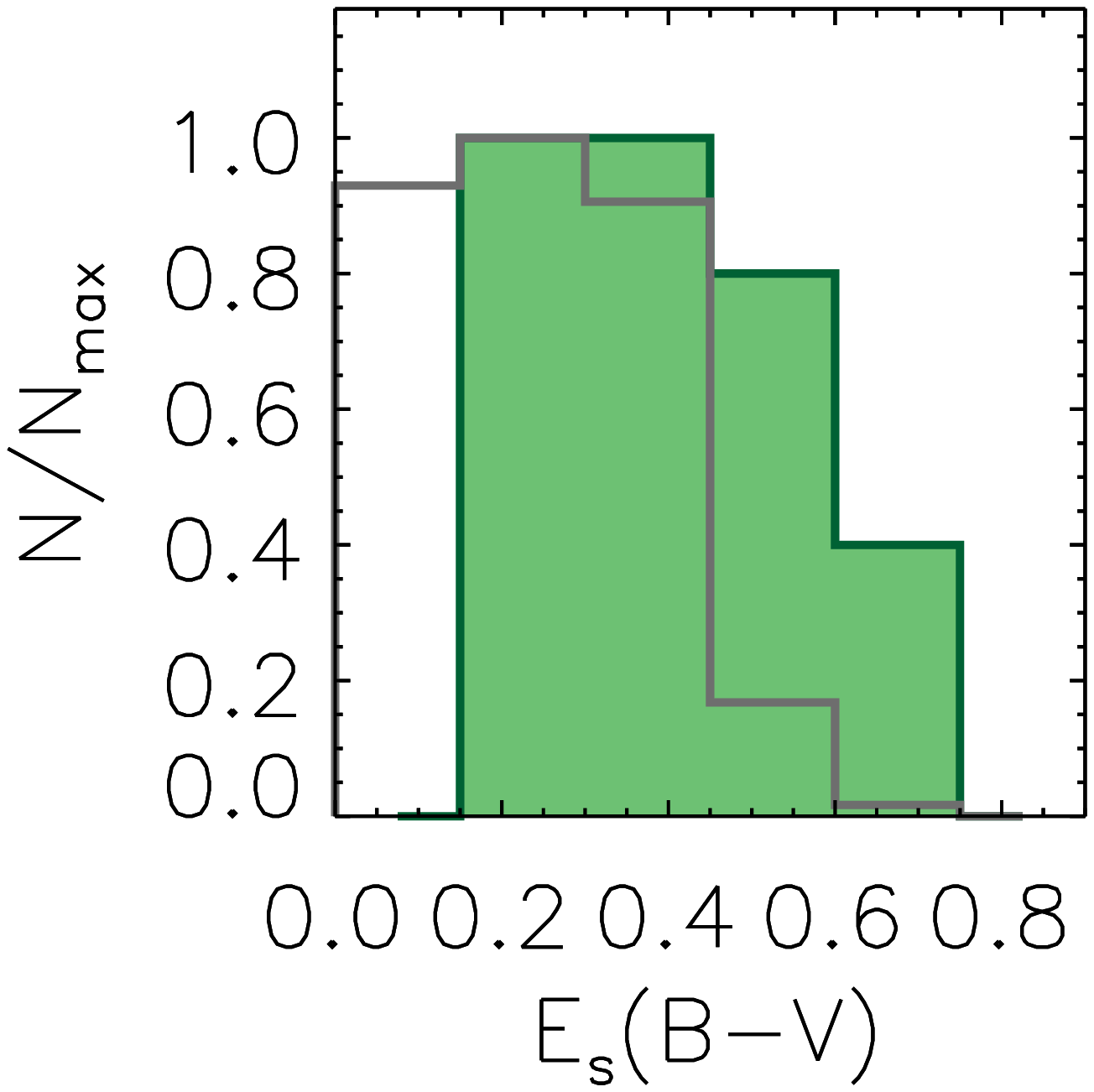}
\includegraphics[width=0.2\textwidth]{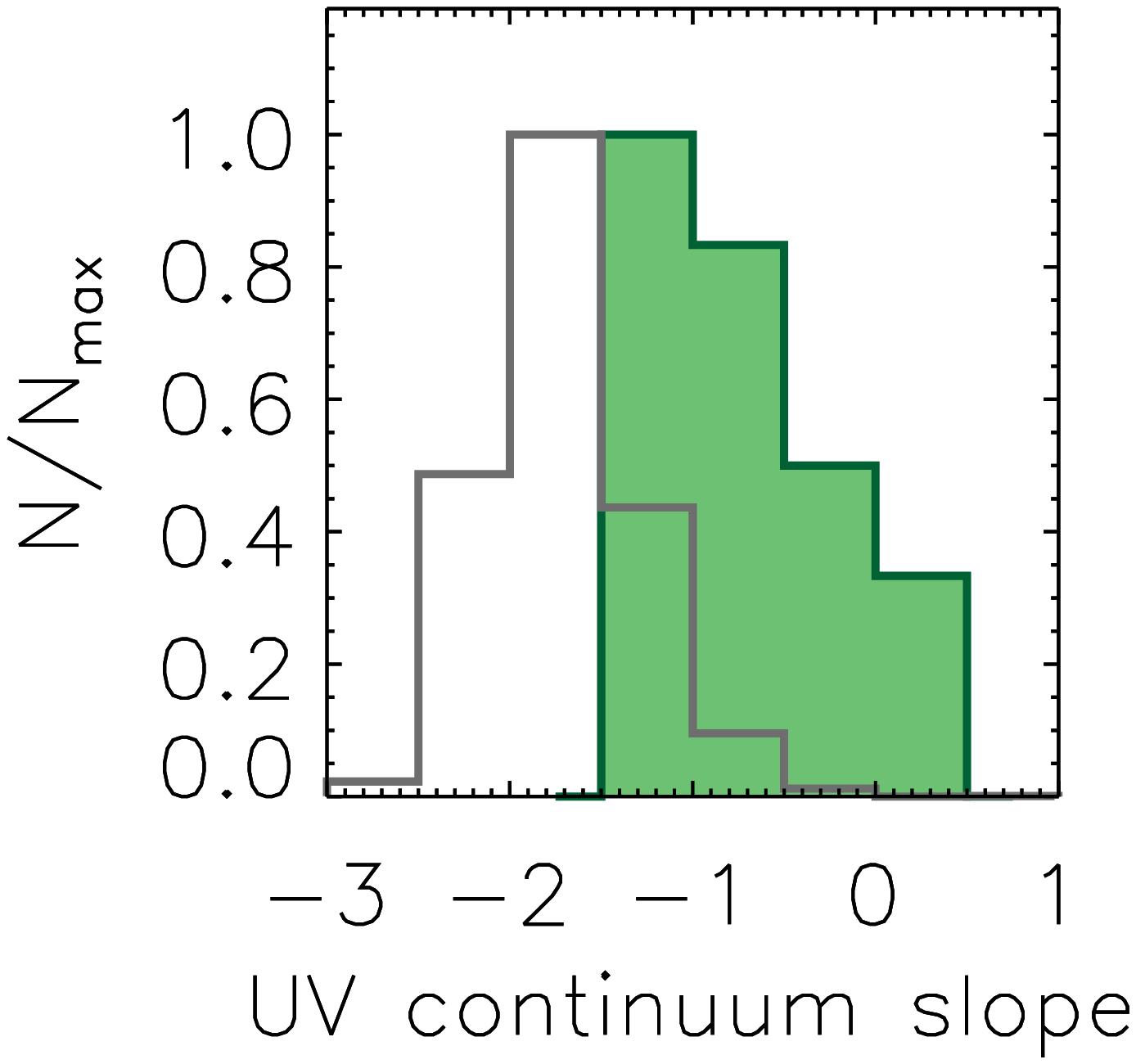}
\includegraphics[width=0.2\textwidth]{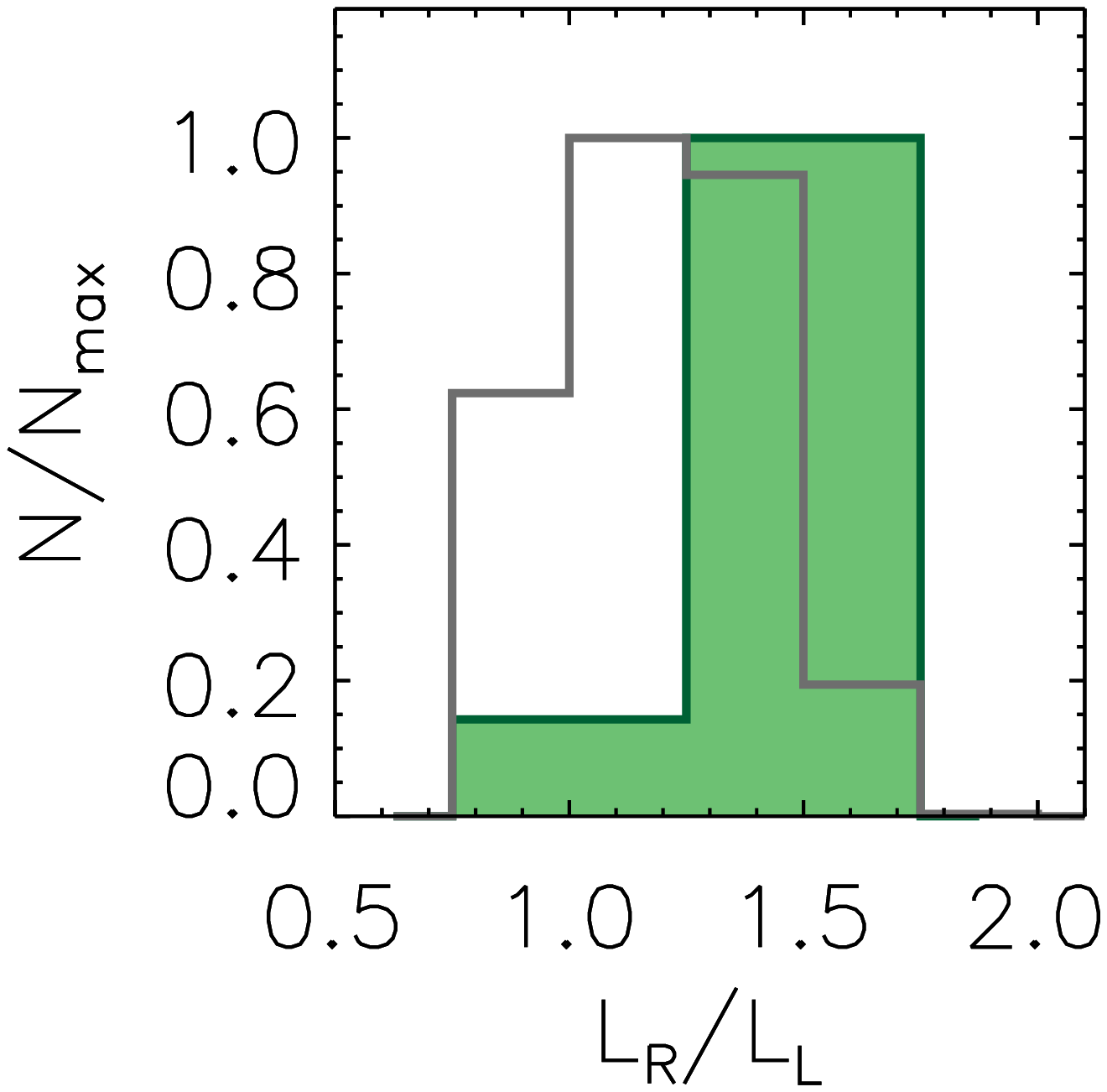}
\caption{Differential SED-derived properties between PACS-detected (shaded green histograms) and PACS-undetected (grey histograms) LBGs at $z \sim $3. The SED-derived properties have been derived by fitting the observed rest-frame UV to near-IR photometry of the galaxies with \cite{Bruzual2003} templates associated to a constant SFR and fixed metallicity of $Z=0.4Z_\odot$. Histograms have been normalized to their maxima in order to clarify the representations.
              }
\label{ppp}
\end{figure*}

In order to understand high-redshift LBGs a deeper study of their FIR spectral energy distribution (SED) is needed. In this work we exploit the deep publicly available FIR data taken in the framework of the GOODS-\emph{Herschel} project \citep{Elbaz2011} in the GOODS-North and GOODS-South fields to increase the sample of FIR-bright PACS/SPIRE-detected LBGs at $z \sim 3$ and analyze in more detail their FIR emission. Additionally, we compare the behavior of FIR-bright LBGs at $z \sim 3$ with their analogues at $z \sim 1$. This letter is structured as follows: the selection of our LBGs and their FIR emission is explained in Section \ref{sample}. In Sections \ref{dd} and \ref{ss} we present and discuss the properties of our PACS-detected galaxies. Finally, the main conclusions of the work are summarized in Section \ref{conclusions}. Throughout this letter we assume a flat universe with $(\Omega_m, \Omega_\Lambda, h_0)=(0.3, 0.7, 0.7)$ and all magnitudes are listed in the AB system \citep{Oke1983}.

\section{FIR emission of LBGs at $z \sim 3$}\label{sample}

We select our LBGs at $z \sim 3$ by employing the classical dropout technique with the broadband filters $U$, $V$, and $I$. Details of the analytical selection criterion can be found in Appendix \ref{sample_selection}. Briefly, a large set of \cite[][hereafter BC03]{Bruzual2003} templates at different redshifts were convolved with the transmission curves of the $U$, $V$, and $I$ filters. Analytically, we select our LBGs through:

\begin{equation}
	U - V \geq 1.71 \times ( V - I ) + 0.57
\end{equation}

\begin{figure*}
\centering
\includegraphics[width=0.4\textwidth]{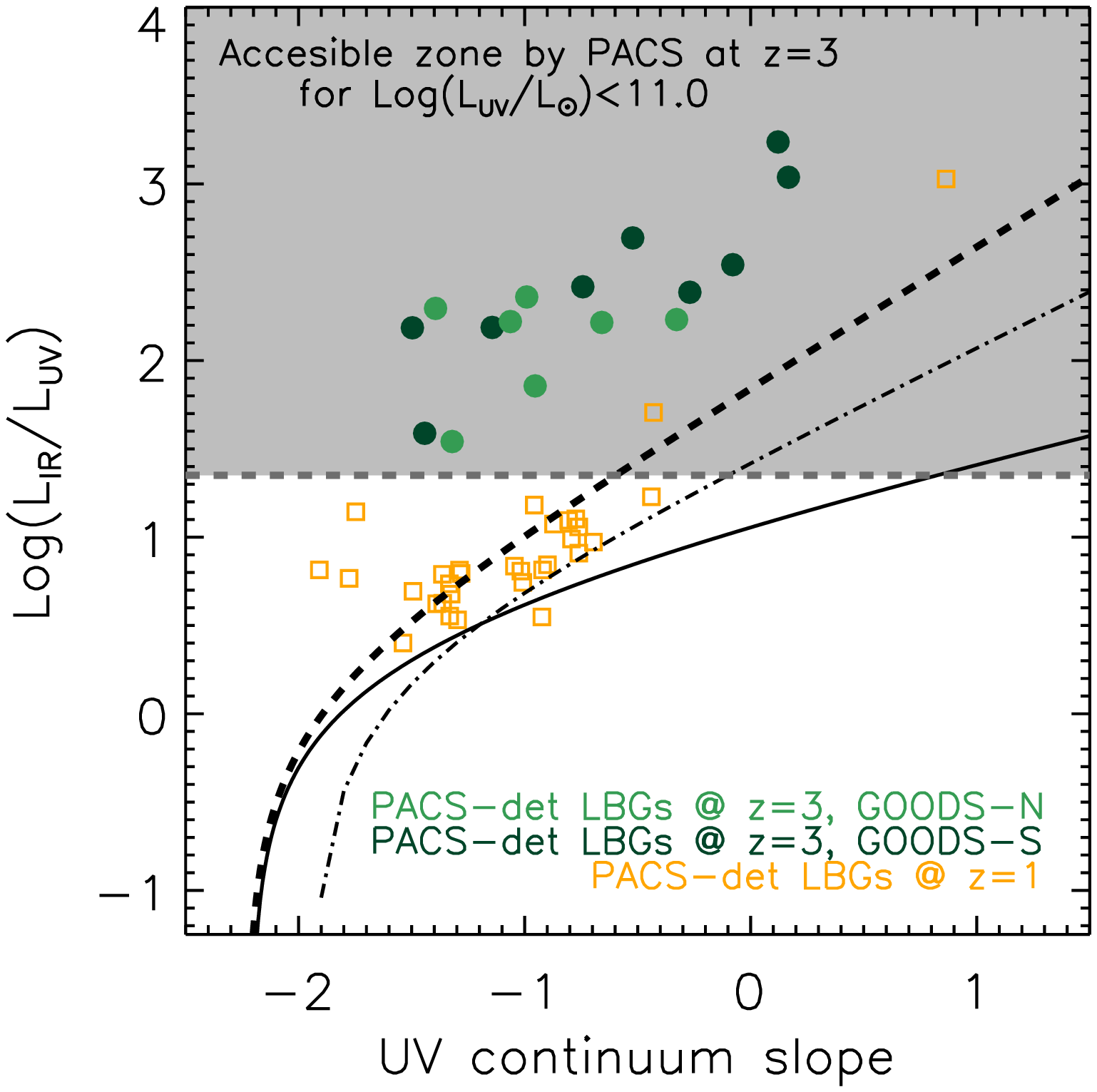}
\includegraphics[width=0.4\textwidth]{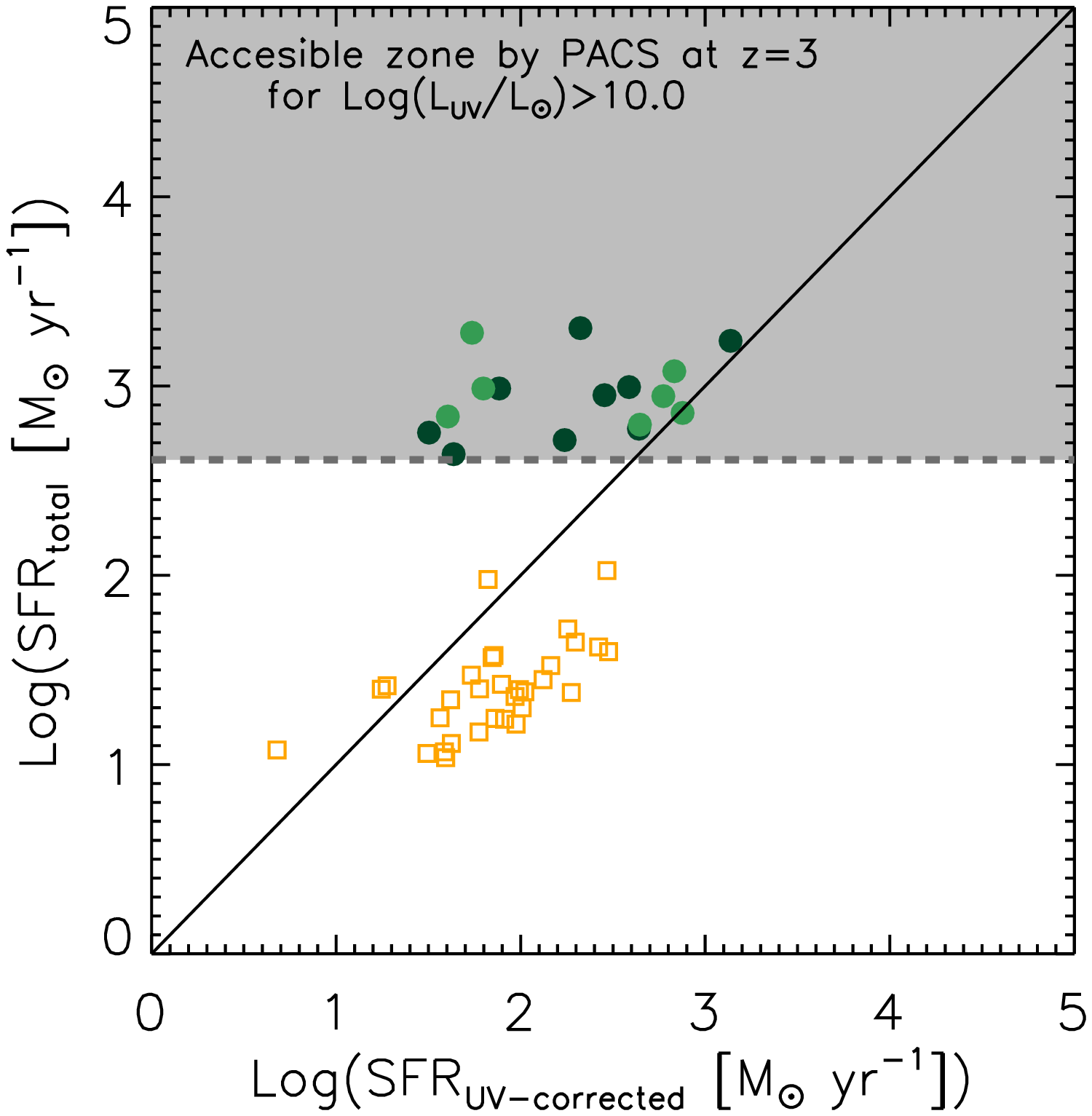}
\includegraphics[width=0.4\textwidth]{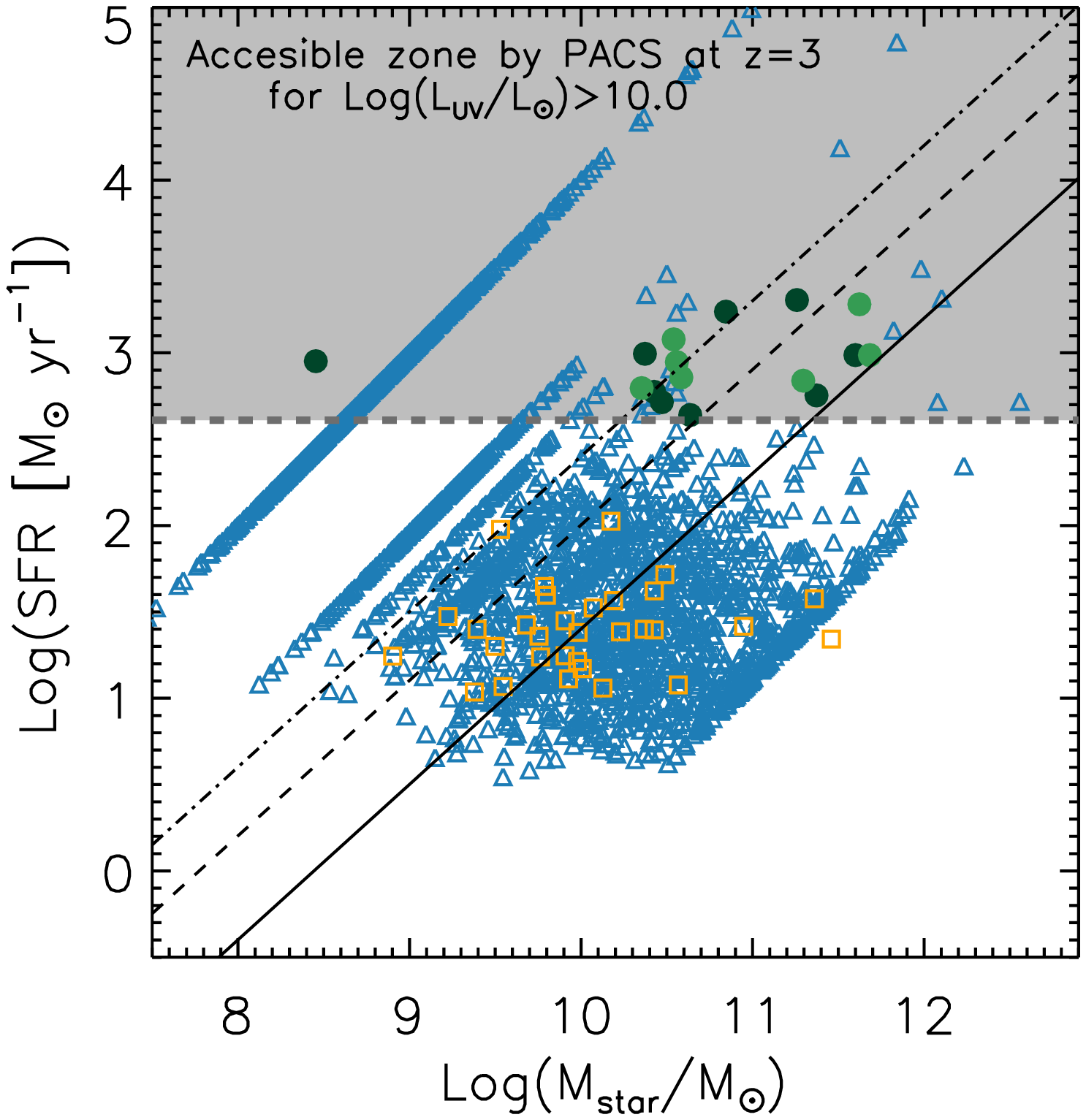}
\includegraphics[width=0.4\textwidth]{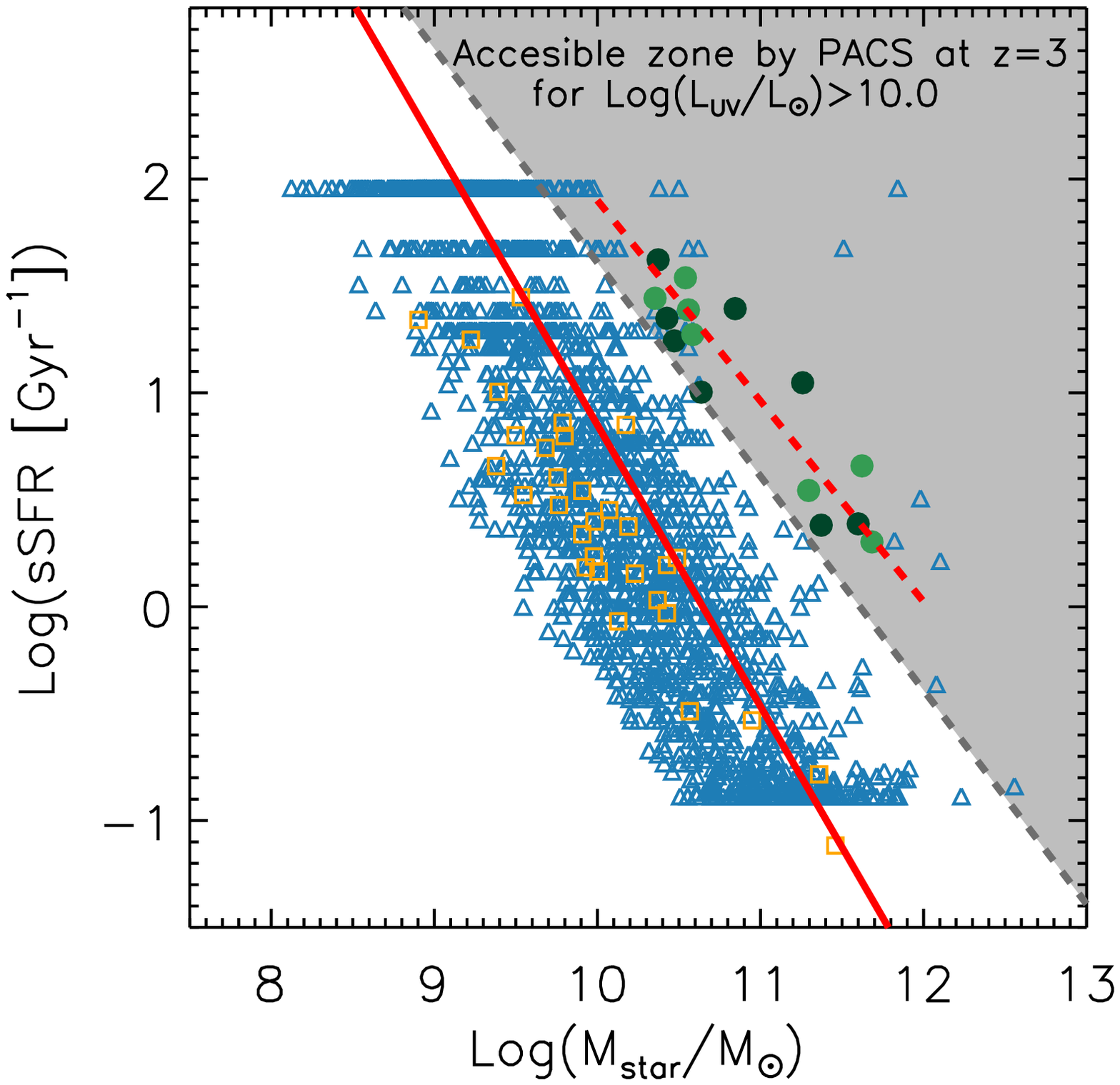}
\caption{FIR-derived properties of our PACS-detected LBGs at $z \sim 3$ (light green dots for galaxies in GOODS-North and dark green dots for galaxies in GOODS-South) and at $z \sim 1$ (orange open squares). The grey shaded zones represent the accessible areas with PACS at $z \sim 3$ for $\log{\left( L_{\rm UV}/L_\odot \right)} > 10.0$ or for $\log{\left( L_{\rm UV}/L_\odot \right)} > 11.0$, which represent the range of the rest-frame UV luminosity for LBGs at $z \sim 3$. In the bottom left and bottom right panels, blue open triangles represent the whole sample of LBGs at $z \sim 3$ in GOODS-South regardless the detection in PACS. The patterns seen in the blue triangles are due to the sampling of the BC03 templates employed to fit the rest-frame UV-to-near-IR SED. In the fourth panel, the red straight line is the linear fit to the whole population of LBGs at $z \sim 3$ and the red dashed line is a linear fit to the PACS-detected LBGs at $z \sim 3$. \emph{Upper left}: Dust attenuation as parametrized by the ratio between the total IR and rest-frame UV luminosities \citep{Buat2005} as a function of the UV continuum slope. The relations of \cite{Meurer1999} for local SB and \cite{Boissier2007} for local normal SF galaxies are represented with dashed and continuous curves, respectively. We also plot the correction of \cite{Takeuchi2012} to the \cite{Meurer1999} relation with a dashed-dotted curve. \emph{Upper right}: UV+IR-derived total SFR against SED-derived dust corrected total SFR. Solid line is the one-to-one relation. \emph{Bottom left}: SFR versus stellar mass plane. The solid line is the MS of galaxies at $z \sim 2$ reported in \cite{Daddi2007}. The dashed and dashed-dotted lines are 4 and 10 times the MS and are employed to separate normal SF galaxies from SB \citep{Rodighiero2011}. \emph{Bottom right}: sSFR versus stelar mass plane. 
              }
\label{figures}
\end{figure*}

With this, we isolate a sample of 2652 and 2537 LBGs candidates in GOODS-North and GOODS-South, respectively. We look for their FIR counterparts by employing PACS and SPIRE publicly available data coming from the GOODS-\emph{Herschel} project. In this step we adopt a matching radius of 2''. We obtain that, among the whole sample of 2652 and 2537 LBGs in GOODS-North and GOODS-South, 7 and 9, respectively, are likely detected in any of the PACS bands. These numbers represent percentages of FIR detection lower than 0.5\%. One LBG in GOODS-South and another in GOODS-North are spectroscopically confirmed to be at $z \sim 3$ (see Figure \ref{steidel}). Additionally, among the 7 PACS-detected LBGs in GOODS-North, 5 of them are also detected in SPIRE-250$\mu$m, SPIRE-350$\mu$m, or SPIRE-500$\mu$m. No SPIRE data is publicly available in GOODS-South by the time of writing of this paper and, therefore, we cannot report on SPIRE-detections in that field. By using 15''$\times$15'' optical and mid-IR cutouts of our PACS-detected sources we have checked that source confusion is unlikely in our sample (see Figure \ref{PACS_detected}). Details on rest-frame UV-to-near-IR and FIR SED fitting can be found in Appendix \ref{sample_selection}. We obtain the total SFR of our PACS-detected LBGs by combining their UV and FIR measurements, $SFR_{\rm total} = SFR_{\rm UV,uncorrected} + SFR_{\rm IR}$, where $SFR_{\rm UV,uncorrected}$ is the SFR associated to the dust-uncorrected rest-frame UV luminosity and $SFR_{\rm IR}$ is the SFR associated to the IR emission of our galaxies. Both terms are calculated by employing the \cite{Kennicutt1998} calibrations.

Figure \ref{ppp} indicates that PACS-detected LBGs are more massive (median $\log{\left( M_*/M_\odot \right)} = 10.7 \pm 0.8$), have higher SED-derived dust attenuation (median $E_s(B-V) = 0.40 \pm 0.15$), stronger Balmer 4000\AA\ break (median $L_R/L_L = 1.44 \pm 0.21$), and redder UV continuum slope (median $\beta = -0.74 \pm 0.56$) than PACS-undetected LBGs. The uncertainties indicate the width of the distributions. There is no significant difference between the age of both populations. Table \ref{properties} compiles some properties of our PACS-detected LBGs at $z \sim 3$. The median redshift of the sources is $z_{\rm phot}=3.1$. It can be seen in Table \ref{properties} that all our PACS-detected LBGs are Ultra luminous IR galaxies (ULIRGs) or Hyper luminous IR galaxies (HyLIRGs). This is a direct consequence of the depth of the FIR data employed: only the FIR-brightest sources are detected. The IR/UV-derived dust attenuation is larger than 5 mag in 1200 \AA\ for most of the galaxies. This indicates that, despite being selected through rest-frame UV color and being UV-bright galaxies, it exists subpopulation of LBGs at $z \sim 3$ that is very dusty. PACS-detected LBGs are extreme SF galaxies with UV+IR-derived total SFR typically higher than 500 $M_{\odot} \, {\rm yr}^{-1}$.

%Masa PACS      10.6366
%Masa no-PACS      9.69991
%LNUV PACS      10.3067
%LNUV no-PACS      10.6009
%dust PACS     0.400000
%dust no-PACS     0.225000
%Balmer break PACS      1.44172
%Balmer break no-PACS      1.16915
%Edad PACS      171.000
%Edad no-PACS      61.0000
%UV-slope PACS    -0.742690
%UV-slope no-PACS     -1.73740

We compare in the next sections the FIR-derived properties of our PACS-detected LBGs at $z \sim 3$ with those of a sample of 32 PACS-detected LBGs at $z \sim 1$ in the GOODS-North and GOODS-South field (see Appendix \ref{sample_selection} for details on the sample selection). There is a fundamental difference between the two populations of FIR-bright LBGs: at $z \sim 1$, all PACS-detected LBGs are LIRGs with  $\log{\left( L_{\rm IR}/L_\odot \right)} < 11.7$ whereas all PACS-detected LBGs at $z \sim 3$ are ULIRGs with $\log{\left( L_{\rm IR}/L_\odot \right)} > 12.4$ or HyLIRGs. LIRGs are not detected at $z \sim 3$ due to the FIR selection bias but ULIRGs at $z \sim 1$ could have been found since the observations are complete down to the expected brightness for ULIRGs in this redshift range. The fact that we do not find any ULIRG-LBGs with $\log{\left( L_{\rm IR}/L_\odot \right)} > 12.4$ at $z \sim 1$ might indicate that the FIR emission of LBGs might have changed with redshift. When analyzing the evolution of the FIR emission of LBGs with redshift, the difference in the surveyed volumes need to be taken into account. Both LBGs at $z \sim 1$ and at $z \sim 3$ are located in the same fields and the comoving volume within the redshift range $2.5 \leq z \leq 4$ is about 5.4 times higher than that in $0.8 \leq z \leq 1.2$. Consequently, the fact that we do not find any ULIRG with $\log{\left( L_{\rm IR}/L_\odot \right)} > 12.4$ in the sample of LBGs at $z \sim 1$ might be also due to the smaller volume surveyed. However, in Oteo et al. (2013b, submitted) we do not find any LBG with an ULIRG nature at $z \sim 1$ either in a volume which is only two times smaller than that surveyed in the present work at $z \sim 3$. Supposing that the density of ULIRGs with $\log{\left( L_{\rm IR}/L_\odot \right)} > 12.4$ is the same at $z \sim 3$ and $z \sim 1$, we should have found 9 galaxies of that type at $z \sim 1$ in Oteo et al. (2013b). The fact that all of the PACS-detected LBGs at $z \sim 1$ have $\log{\left( L_{\rm IR}/L_\odot \right)} < 11.7$ reinforces the idea of an evolution of the FIR emission of LGBs with redshift. This would be in agreement with the evolution of the evolution of the SFR density (SFRD) of the universe between $z \sim 1$ and $z \sim 3$ \citep[see for example][]{Hopkins2006}. The higher values of the SFRD at $z \sim 3$ might favor the appearance of such as extreme LBGs at high redshift.

%\section{Results}\label{results}

\section{Dust-correction factors}\label{dd}

The upper left panel of Figure \ref{figures} indicates that, for each value of the UV continuum slope ($\beta$), the dust attenuation parametrized by $IRX=\log{\left( L_{\rm IR} / L_{\rm UV} \right)}$ of the PACS-detected LBGs at $z \sim 3$ is much higher than that predicted by the star-burst (SB) relations of \cite{Meurer1999} and \cite{Takeuchi2012}. This is in agreement with FIR-bright galaxies at low redshift \citep{Goldader2002}. Therefore, there is a population of LBGs at $z \sim 3$ for which the SB relations fails at recovering their dust attenuation from their UV continuum slopes. Specifically, this method would produce underestimated results. For a given UV continuum slope, the dustiest LBGs at $z \sim 1$ have lower dust attenuation than the dustiest LBGs at $z \sim 3$. For each $\beta$, LBGs at $z \sim 1$ as dusty as those at $z \sim 3$, could have been found since the observations are complete. The absence of this kind of sources suggests that the upper envelope of the locus of LBGs in a IRX-$\beta$ diagram might have changed with redshift. Interestingly, despite PACS-detected LBGs at $z \sim 3$ are dustier than those at $z \sim 1$, the UV continuum slopes of both populations span within a similar range.
 
The upper right panel of Figure \ref{figures} indicates that the SED-derived dust attenuation of our PACS-detected LBGs at $z \sim 3$ cannot be used to recover their UV+IR-derived total SFR. The SED-derived total SFR is underestimated for most of the galaxies. This might be a consequence of a clumpy or patchy geometry of the dust regions, and the UV-luminous part and IR-luminous part are different in each galaxy. The reported underestimation is in agreement with previous results suggesting that SED-fitting procedures fails for estimating the total SFR in dust-obscured high-redshift galaxies \citep{Oteo2012a,Wuyts2011_SED}. At $z \sim 1$, instead of an underestimation, the SED-derived dust attenuation overestimates the total SFR for LBGs. The SED-derived total SFR of the PACS-detected LBGs at $z \sim 1$ and $z \sim 3$ spans over a similar range, whereas the UV+IR-derived SFR is much higher for PACS-detected LBGs at higher redshifts. The total SFR might be also recovered from the UV continuum slope by applying, for example, the \cite{Meurer1999} law or the correction given by \cite{Takeuchi2012} \citep{Oteo2013a,Bouwens2009,Bouwens2012,Castellano2012}. In this case, since our PACS-detected LBGs are well above the SB relations, this technique would also underestimate the real dust attenuation, and thus the total SFR, of the studied galaxies. Therefore, we conclude that the only way to obtain reliable values of the dust attenuation and total SFR of our PACS-detected LBGs at $z \sim 3$ is using their UV and FIR emission. This highlights the importance of FIR data when studying high-redshift sources. 

\section{SFR and stellar mass}\label{ss}

The location of our PACS-detected LBGs at $z \sim 3$ in an SFR versus stellar mass diagram is shown in the bottom left panel of Figure \ref{figures}. For comparison, we also plot the location of the whole population of LBGs at $z \sim 3$ in GOODS-North regardless their detection in the FIR. For a given stellar mass, PACS-detected LBGs have higher total SFR than those PACS-undetected. It should be remarked that the total SFR for PACS-detected and PACS-undetected galaxies has been calculated differently. For the former, the total SFR is derived from their direct UV and FIR emission. For the latter, the total SFR has been obtained by correcting the rest-frame UV luminosity with the best-fitted value of the SED-derived dust attenuation, as explained in Section \ref{sample}. We also show in that panel the SFR-stellar mass relation of the main sequence (MS) at $z \sim 2$ reported in \cite{Daddi2007}. Following \cite{Rodighiero2011} we represent 4 and 10 times the MS of \cite{Daddi2007} as a reference for the definition of SB galaxies (see also \cite{Elbaz2011}). At $z \sim 3$, the low number of FIR detections of SF galaxies has prevented the definition of an MS. Thus we have to base our results on the MS at $z \sim 2$ without forgetting that there are evidences of an evolution of the MS with redshift \citep{Elbaz2011}. It can be seen that most PACS-detected LBGs at $z \sim 3$ are located above 4 times the MS and, therefore, have a SB nature. The FIR-brightest LBGs at $z \sim 3$ have higher values of the total SFR for a given stellar mass than the FIR-brightest LBGs at $z \sim 1$. This indicates an evolution of the high-SFR tail of LBGs with redshift.

The sSFR versus stellar mass for our PACS-detected LBGs at $z \sim 3$ is presented in the bottom right panel of Figure \ref{figures}. It can be seen that PACS-detected LBGs at $z \sim 3$ follow a linear relation as a direct consequence of the limiting FIR fluxes of the observations employed: only very FIR-bright galaxies with high total SFRs are detected. For each stellar mass, PACS-detected LBGs at $z \sim 3$ have higher values of the sSFR than those PACS-undetected. For a given stellar mass, the FIR-brightest LBGs at $z \sim 1$ have lower values of the sSFR than those at $z \sim 3$. This again indicates an evolution of the FIR emission of LBGs with redshift, at least in the FIR-brightest side.

\section{Conclusions}\label{conclusions}

In this work we have found a sample of 16 FIR-bright LBGs at $2.5 \lesssim z \lesssim 4.0$ in GOODS-North and GOODS-South fields. These galaxies are individually detected in PACS-100$\mu$m or PACS-160$\mu$m, probing directly their dust emission. These detections allow us to determine their total IR luminosities, dust attenuation, and total SFR without the uncertainties that introduce the traditionally used SED-fitting technique  with BC03 templates. The dust attenuation and total SFR of these objects cannot be recovered from the dust correction factors obtained with their UV slope or their SED-derived dust attenuation since both methods underestimate the results. The only way of obtaining accurate results is employing UV and FIR data, highlighting the importance of these wavelengths for the studies of high-redshift sources. Comparing with a sample of PACS-detected LBGs at $z \sim 1$ we find evidence that the FIR emission of LBGs might have evolved with redshift: the dustiest LBGs at $z \sim 3$ have more  prominent FIR emission, have higher dust attenuation for a given UV slope, and have higher total SFR for a given stellar mass than the dustiest LBGs at $z \sim 1$.

\begin{acknowledgements}

The authors would like to thank to the referee, Tsutomu T. Takeuchi, for the useful comments provided, that have improved the presentation of the results reported in this letter. This research has been supported by the Spanish Ministerio de Econom\'ia y Competitividad (MINECO) under the grant AYA2011-29517-C03-01. Some/all of the data presented in this paper were obtained from the Multimission Archive at the Space Telescope Science Institute (MAST). STScI is operated by the Association of Universities for Research in Astronomy, Inc., under NASA contract NAS5-26555. Support for MAST for non-HST data is provided by the NASA Office of Space Science via grant NNX09AF08G and by other grants and contracts. Based on observations made with the European Southern Observatory telescopes obtained from the ESO/ST-ECF Science Archive Facility. Based on zCOSMOS observations carried out using the Very Large Telescope at the ESO Paranal Observatory under Programme ID: LP175.A-0839. Based on observations made with ESO Telescopes at the La Silla or Paranal Observatories under programme ID 171.A-3045.

\end{acknowledgements}

\bibliographystyle{aa}
\bibliography{ioteo_biblio}

\Online

\onecolumn

\begin{appendix} %First online appendix
\section{Sample selection and SED fitting}\label{sample_selection}

\begin{figure*}
\centering
\includegraphics[width=0.4\textwidth]{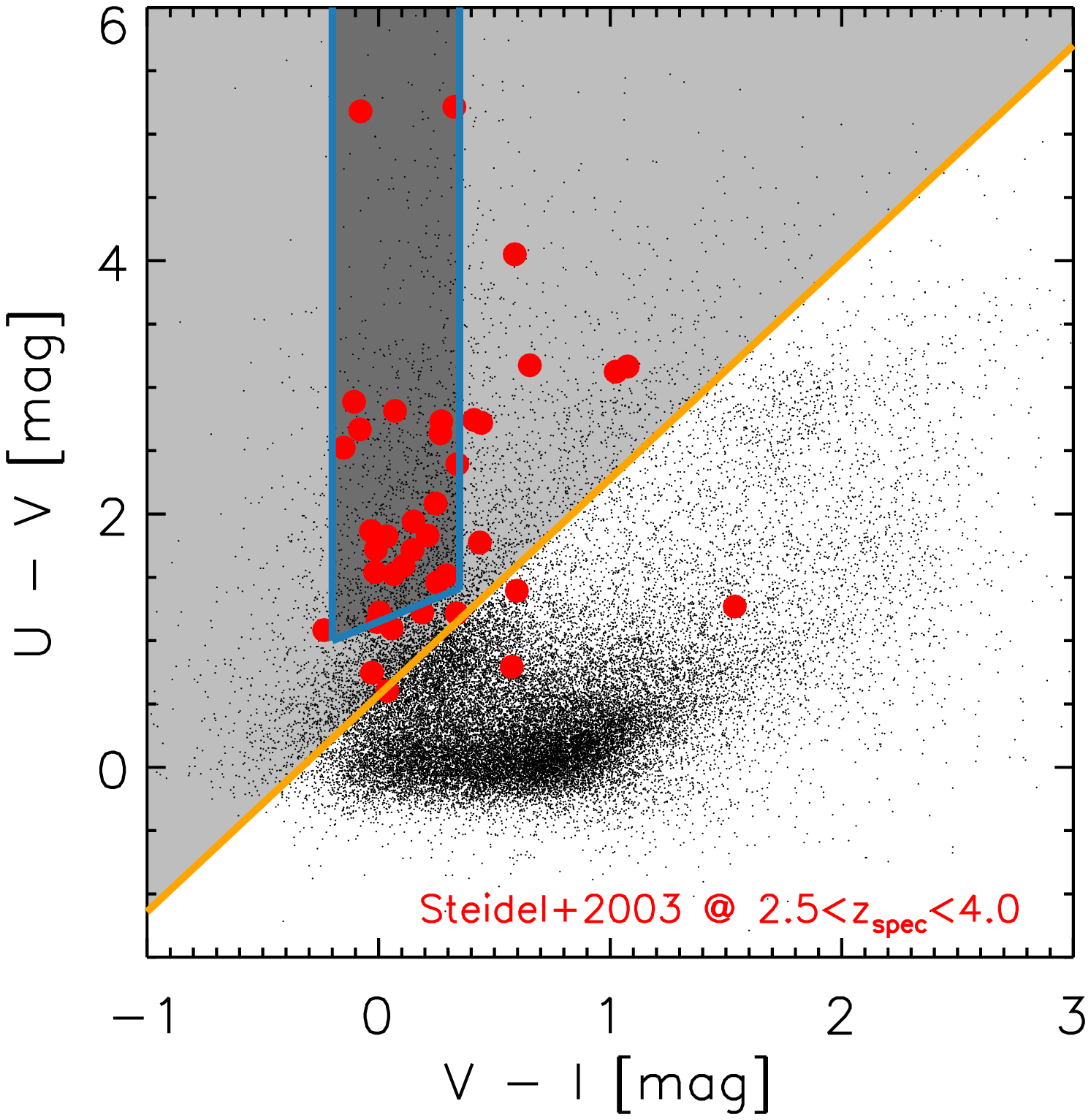}
\includegraphics[width=0.4\textwidth]{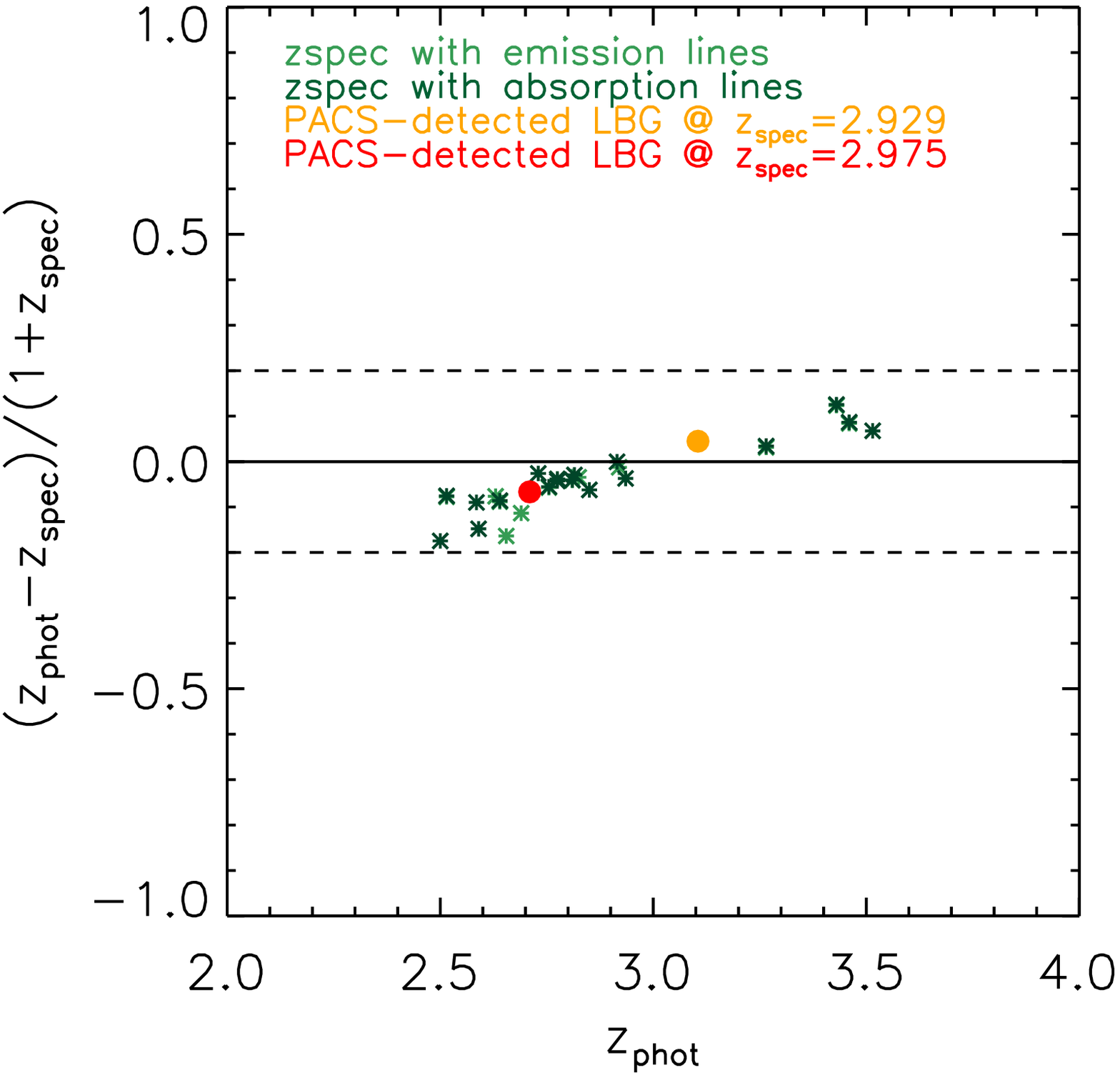}
\caption{\emph{Left plot}: location of the LBGs of \cite{Steidel2003} with confirmed spectroscopic redshifts (red dots) in the color-color diagram employed to select the LBGs studied in this work. The light grey shaded zone represents our selection window for LBGs at $z \sim 3$ (see Equation \ref{criterion}). Black open squares are the complete sample of galaxies in the \cite{Capak2004} photometric catalog. For comparison, we also show the selection window for LBGs of \cite{Pentericci2010} with a dark grey shaded zone. \emph{Right plot}: Photometric redshift accuracy for our LBGs at $z \sim 3$. Only galaxies with confirmed spectroscopic redshift via emission lines (light green symbols) or absorption lines (dark green symbols) in \cite{Steidel2003} are included. The PACS-detected LBG spectroscopically confirmed to be at $z = 2.929$ in \cite{Steidel2003} is indicated with a orange dot. The PACS-detected LBG spectroscopically confirmed to be at $z = 2.975$ in \cite{Popesso2009} and \cite{Balestra2010} is presented with a red dot.
              }
\label{steidel}
\end{figure*}

Our LBGs are selected through the classical dropout technique with the $U$, $V$, and $I$ filters. By convolving the transmission curves of those filters with a large set of \cite{Bruzual2003} (hereafter BC03) templates associated to different physical properties and redshifts ($0 \leq z \leq 6$) \citep[see for example][]{Oteo2013a}, we derive the following analytical selection criterion:

\begin{equation}\label{criterion}
U - V \geq 1.71 \times (V - I) +0.57
\end{equation}

The multi-wavelength photometric information from the optical to mid-IR is taken from \cite{Santini2009}, \cite{Capak2004}, and \cite{Wang2010}. We focus our work on the GOODS-North and GOODS-South fields since these are the two fields covered with the GOODS-\emph{Herschel} FIR observations. The criterion shown in Equation \ref{criterion} is less restrictive than others employed in previous studies to look for LBGs at $z \sim 3$ \citep{Grazian2007,Pentericci2010}. It can be seen in the left panel of Figure \ref{steidel} that our selection window covers a wider area in the color-color diagram than that represented by the \cite{Pentericci2010} criterion. We also show in the left panel of Figure \ref{steidel} the location of the spectroscopically confirmed LBGs at $z \sim 3$ of \cite{Steidel2003} in the GOODS-North field. To do that we have used the same photometric information than that employed to look for our LBGs, i.e. the photometric catalog of \cite{Capak2004}. Most of the spectroscopically confirmed LBGs at $z \sim 3$ of \cite{Steidel2003} are within our selection window and, therefore, we conclude that Equation \ref{criterion} truly segregate LBGs at $z \sim 3$. In order to clean the sample from lower-redshift interlopers, we only select those galaxies whose photometric redshifts (see next paragraph) are within $2.5 \leq z_{\rm phot} \leq 4$. We discard AGNs by ruling out sources with X-ray emission. In total, we segregate 2652 and 2537 LBGs candidates in GOODS-North and GOODS-South, respectively. 

%\subsection{SED fitting}

Photometric redshifts, age, dust attenuation, and stellar mass of our LBGs are obtained simultaneously by fitting their $U$ to IRAC-8$\mu$m with a large set of BC03 templates associated to different physical properties of galaxies. LBGs at $z \sim 3$ are not detected in the UV and the MIPS-24$\mu$m have a significant contribution of dust emission features which are not considered in the elaboration of the BC03 templates. For these reasons, these wavelengths are not employed in the SED fits. We build the BC03 templates by using the software \verb+GALAXEV+. In this process we adopt a \cite{Salpeter1955} initial mass function (IMF) distributing stars from 0.1 to 100 M$_\odot$ and select a constant value for metallicity of Z=0.4Z$_{\odot}$. We consider values of age from 1 Myr to 7 Gyr in steps of 10 Myr from 1 Myr to 1 Gyr and in steps of 100 Myr from 1 Gyr to 7 Gyr. Dust attenuation is included in the templates via the \cite{Calzetti2000} law and parametrized through the color excess in the stellar continuum, $E_s(B-V)$. We select values for $E_s(B-V)$ ranging from 0 to 0.7 in steps of 0.05. We include intergalactic medium absorption adopting the\cite{Madau1995} prescription. Regarding SFR, we adopt time-constant models. In this case, different values of the SFR does not change the shape of the templates, and the SFR can be obtained by using the \cite{Kennicutt1998} calibration: 

\begin{equation}\label{SFR_UV}
\textrm{SFR}_{UV,uncorrected}[M_{\odot}\textrm{yr}^{-1}] = 1.4 \times 10^{-28}L_{1500}
\end{equation}

\noindent where $L_{1500}$ is the rest-frame UV luminosity in 1500\AA. The $L_{1500}$ is obtained for each galaxy by convolving its best-fitted template with a top-hat filter centered in rest-frame 1500 \AA. It should be noted that, throughout the work, we distinguish between L$_{UV}$ defined in a $\nu$L$_{\nu}$ way and L$_{1500}$ considered in L$_{\nu}$ units. The SFR derived from Equation \ref{SFR_UV} is uncorrected for the attenuation that dust produces in the SED of galaxies. In order to obtain an estimation of the dust-corrected total SFR we have to introduce into Equation \ref{SFR_UV} the dust-corrected $L_{1500}$. It is obtained from $L_{1500}$ by multiplying it by the dust correction factor 10$^{0.4A_{1500}}$, where $A_{1500}$ is the dust attenuation in 1500\AA. The values of $A_{1500}$ are obtained from the SED-derived $E_s(B-V)$ assuming the \cite{Calzetti2000} law. Throughout the work, the total SFR calculated in this way will be called \emph{SED-derived total SFR}, in contraposition with the more accurate \emph{UV+IR-derived total SFR} obtained with the direct emissions in the UV and FIR (see later in the text). Once both age and dust-corrected total SFR are known for each source, and according to the assumed time-independent SFH, the stellar mass can be obtained from the product of both quantities.

%The rest-frame UV luminosity, $L_{\rm UV}$, of each galaxy is obtained by integrating each best-fitted template with a top-hat filter centered in rest-frame 1500 \AA. The SED-derived total SFR is then obtained by multiplying $L_{\rm UV}$ by the dust correction factor 10$^{0.4A_{1500}}$ and applying the classical \cite{Kennicutt1998} calibration. The parameter $A_{1500}$ is the SED-derived dust attenuation in 1500\AA\ calculated from the SED-derived $E_s(B-V)$ assuming the \cite{Calzetti2000} law. 

We define the amplitude of the Balmer 4000 \AA\ break as the ratio between the rest-frame 4500 and 3500 \AA\ luminosities. These are also calculated by convolving each best-fitted template with top-hat filters centered in 4500 and 3500 \AA, respectively. The UV continuum slopes are calculated by fitting the UV continuum of each best-fitted template with a power law function \citep[see for example][]{Oteo2013a,Finkelstein2012} between rest-frame 1300 and 3000 \AA. We have not included the contribution of emission lines in the SED fits \citep{Zackrisson2008,Schaerer2009,Schaerer2010,Schaerer2011,Schaerer2013,deBarros2013} since all our PACS-detected LBGs have $m_{\rm 3.6\mu m} - m_{\rm 4.5 \mu m} > 0$ and, according to \cite{deBarros2013}, the contribution of emission lines in their SED-derived parameters is not expected to be significant. The right panel of Figure \ref{steidel} shows the accuracy of the photometric redshifts of our LBGs, that we define as $\sigma_{\Delta z} =  |z_{phot} - z_{spec}|/(1+z_{spec} )$ \citep[see for example][]{Oteo2013a}, is lower than 0.2 for all the sources. These values are comparable with those obtained at lower redshifts \citep{Haberzettl2012} and it is enough for our purposes.

\begin{figure*}
\centering
\includegraphics[width=0.23\textwidth]{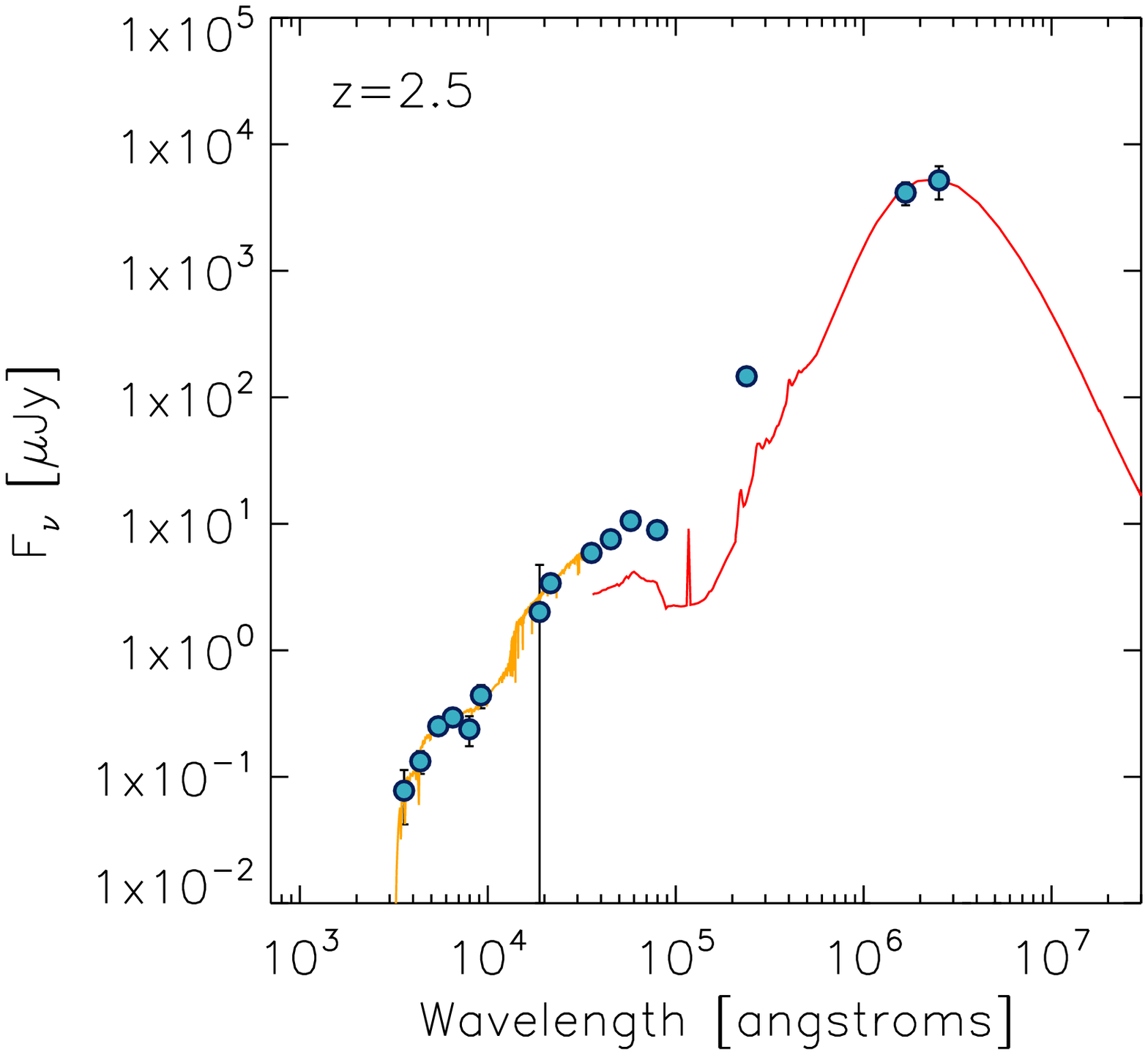}
\includegraphics[width=0.23\textwidth]{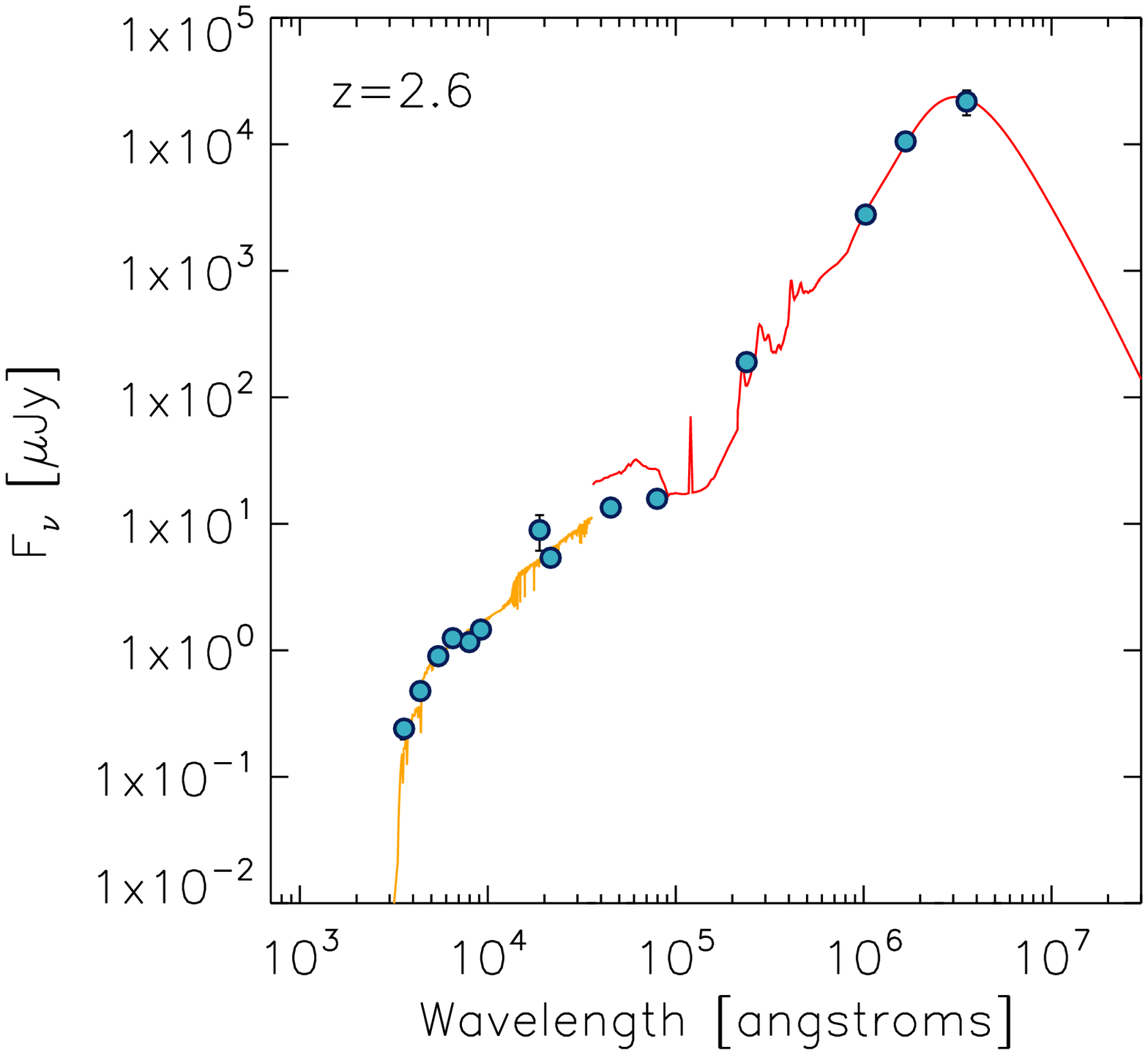}
\includegraphics[width=0.23\textwidth]{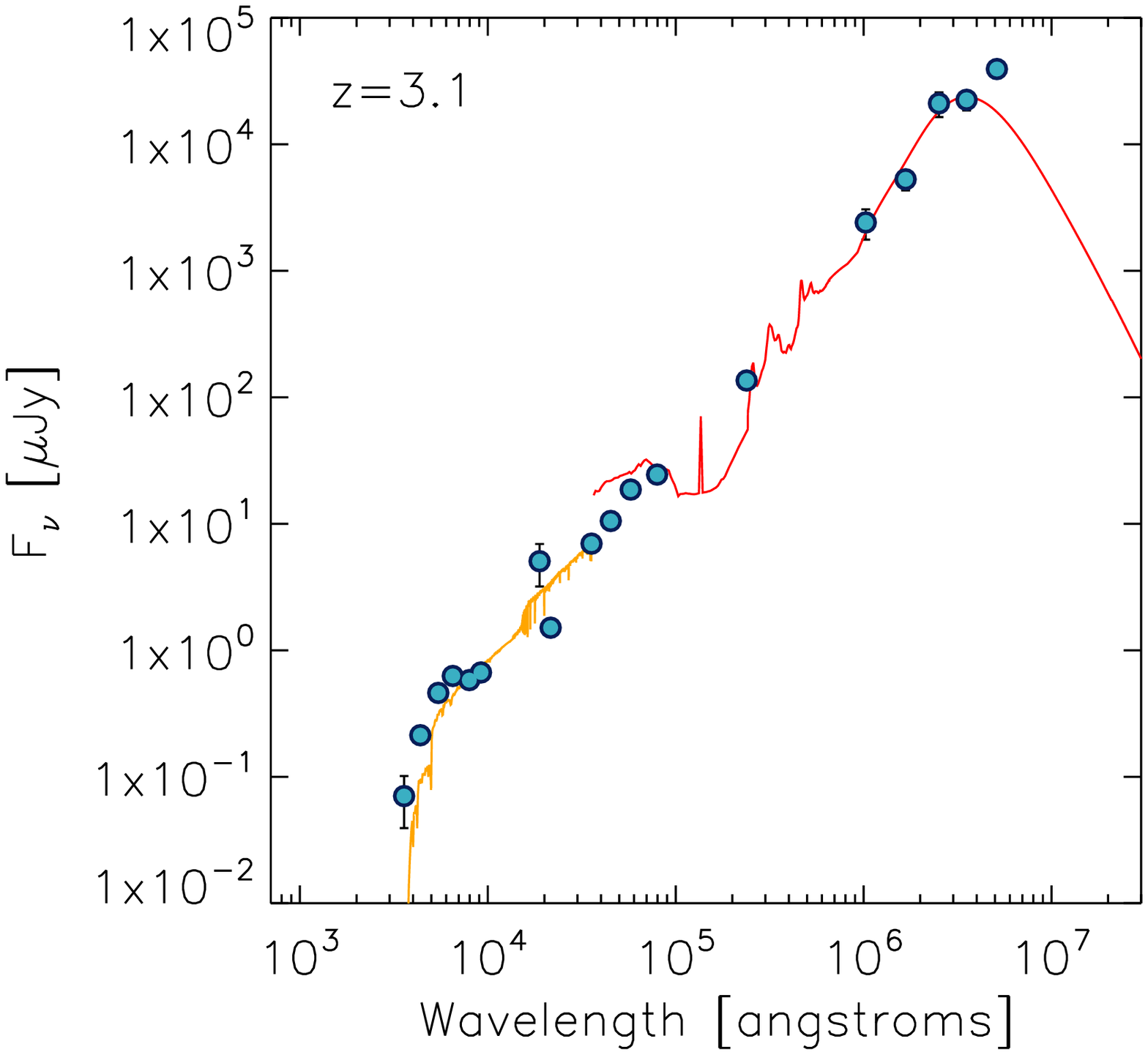}\\
\includegraphics[width=0.23\textwidth]{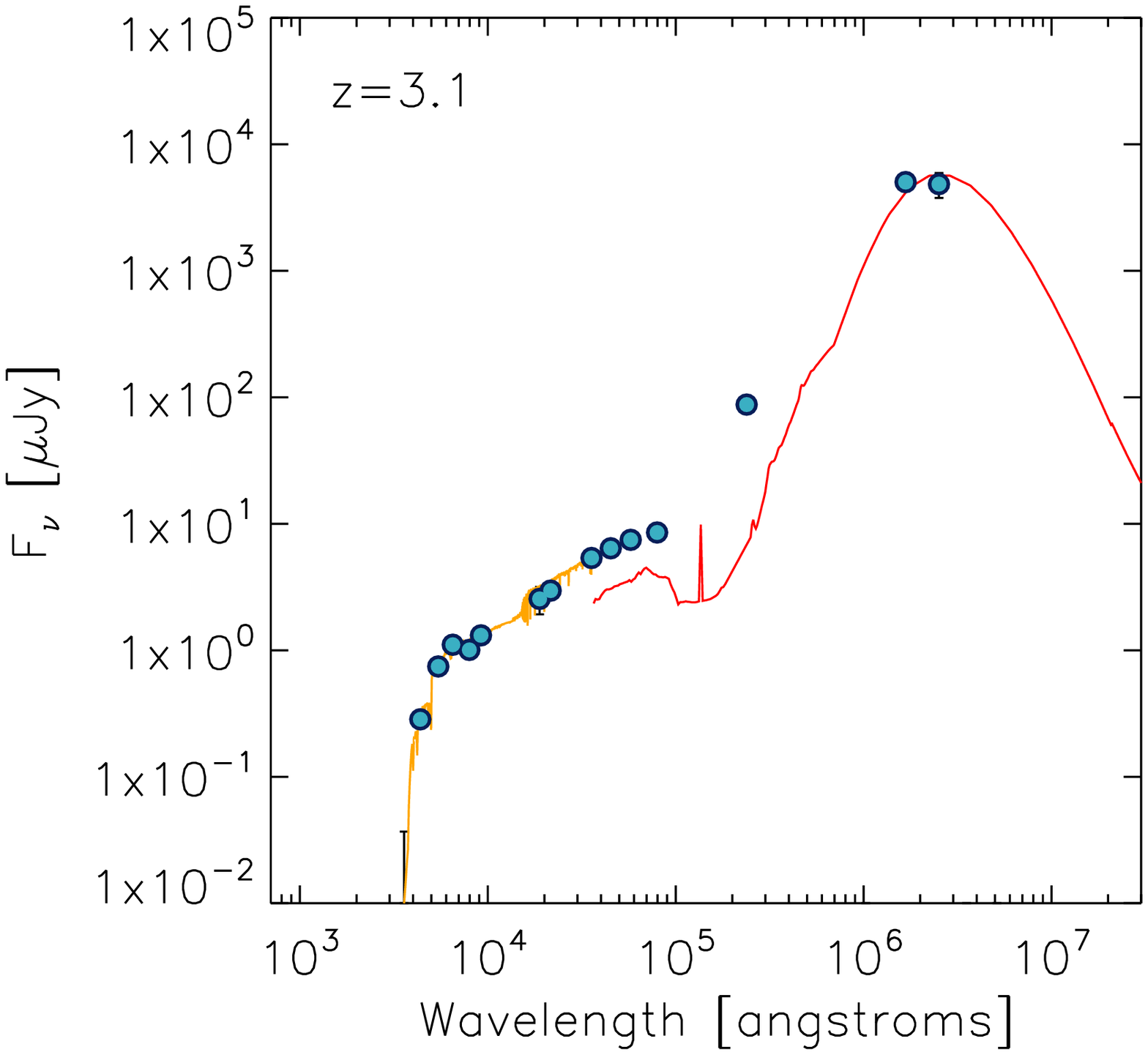}
\includegraphics[width=0.23\textwidth]{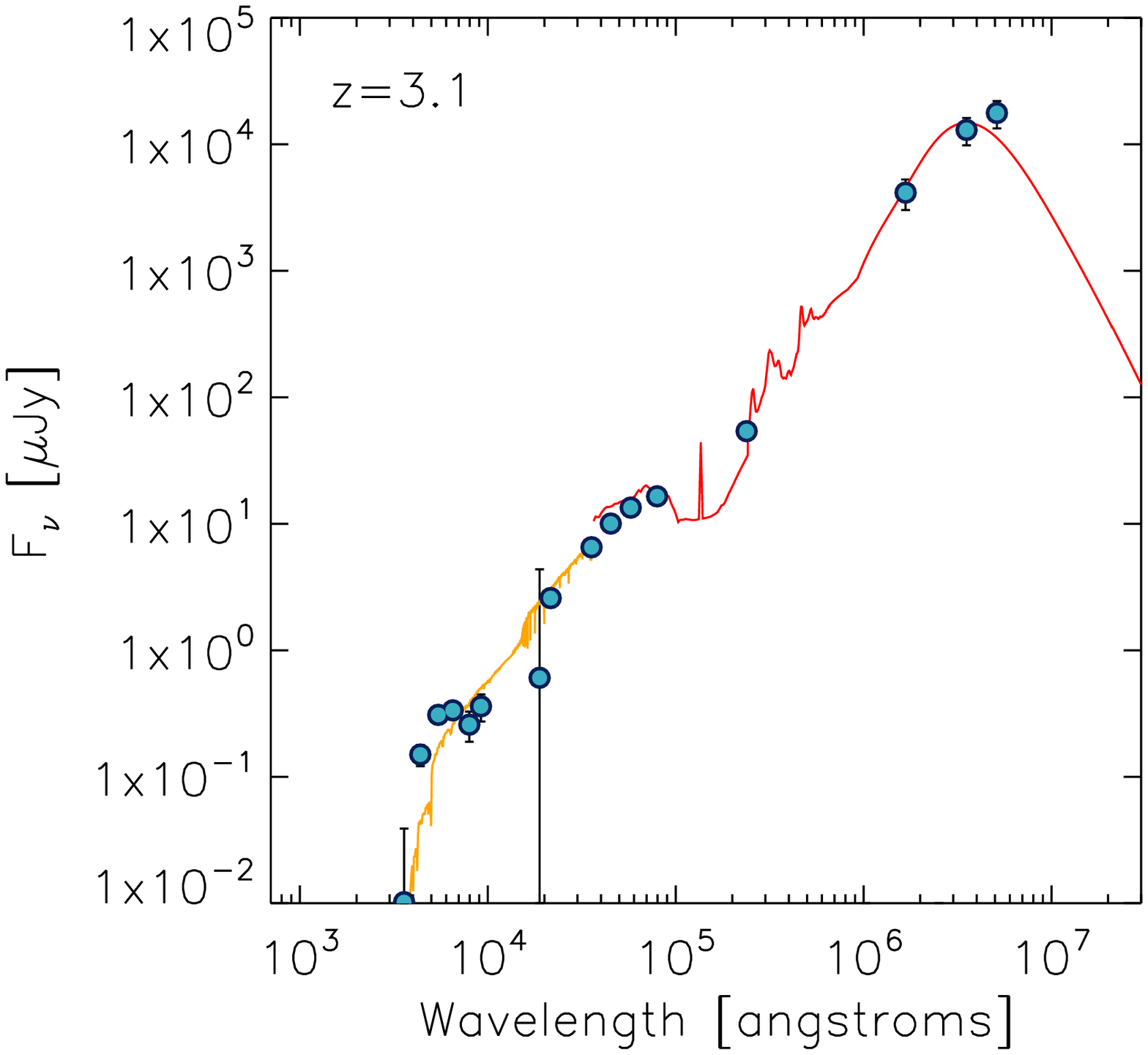}
\includegraphics[width=0.23\textwidth]{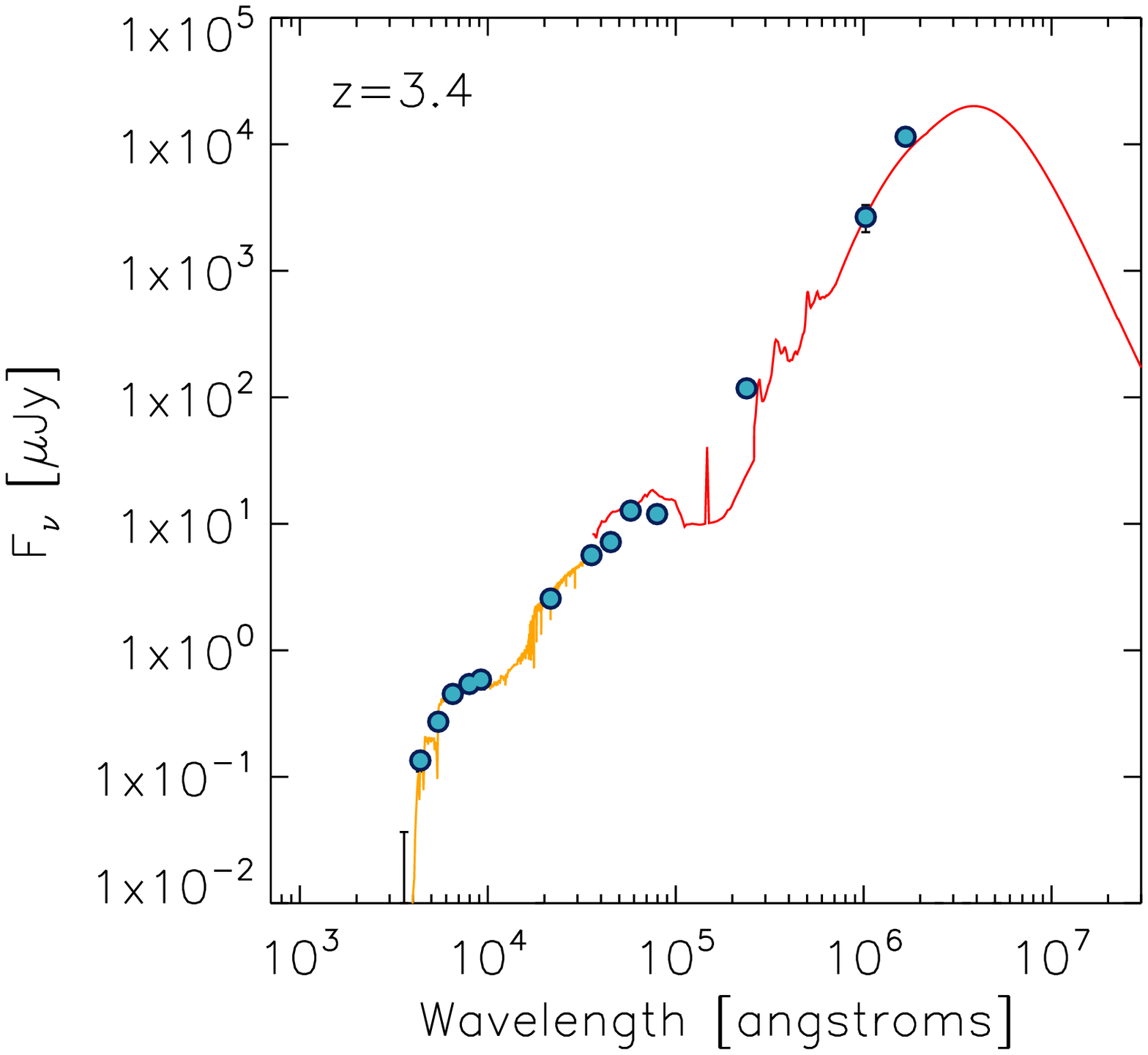}
\includegraphics[width=0.23\textwidth]{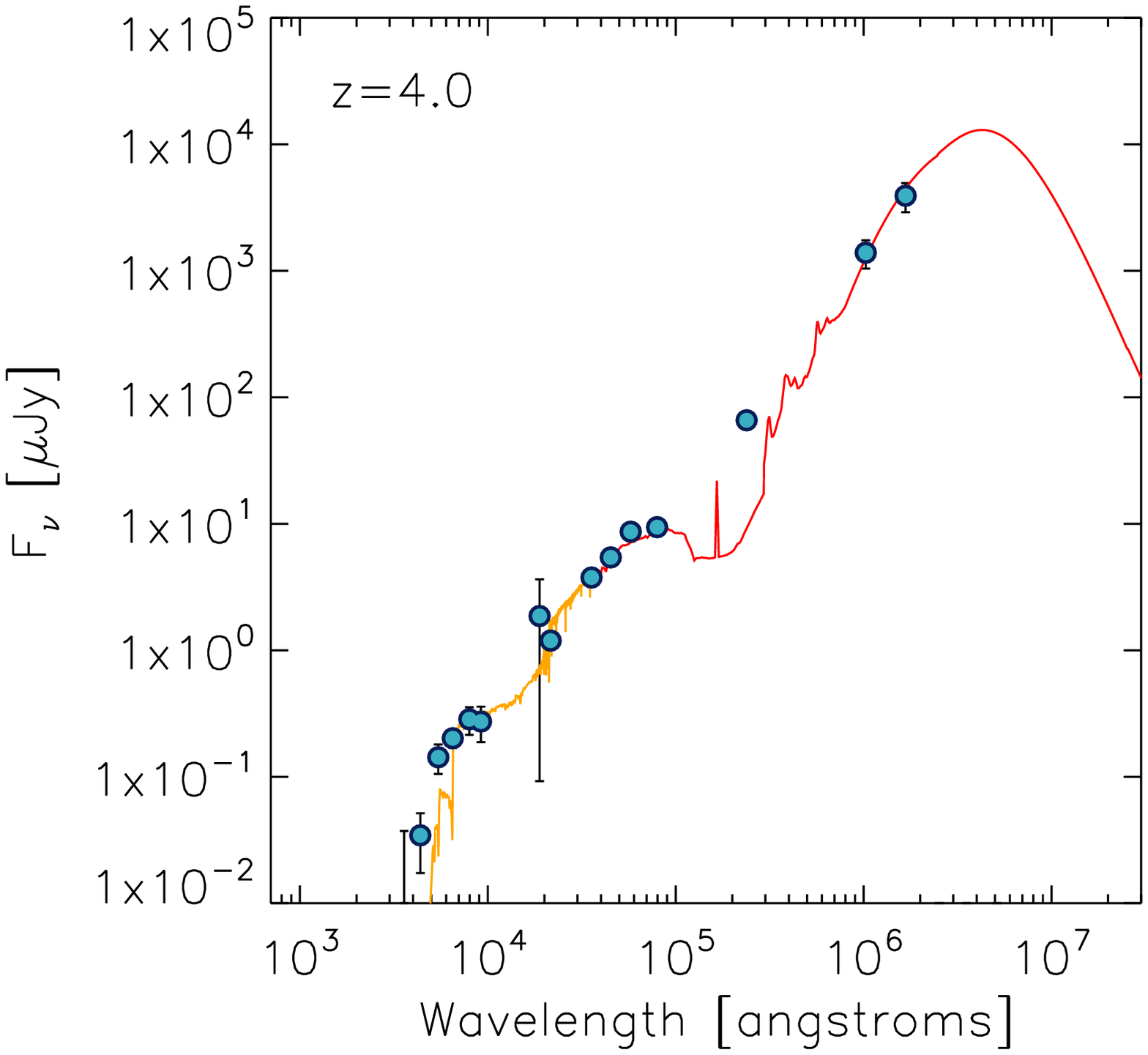}
\caption{Rest-frame UV-to-FIR SEDs of the 7 PACS-detected LBGs at $z \sim 3$ in the GOODS-North fields. These plots represent the typical SED of the studied galaxies. Orange curves represent the best-fitted \cite{Bruzual2003} templates to the U-band to IRAC-8$\mu$m fluxes. Red curves represent the best-fitted \cite{Chary2001} templates to the MIPS-24$\mu$m to PACS and SPIRE fluxes. The photometric redshift of each source is indicated in each panel.
              }
\label{SED_PACS}
\end{figure*}

For all our PACS-detected LBGs we fit \cite{Chary2001} templates to their IRAC-8$\mu$m, MIPS-24$\mu$m, PACS, and SPIRE fluxes (when available). As an example of the typical FIR SED-fitting results, we show in Figure \ref{SED_PACS} the rest-frame UV-to-FIR SEDs of the 7 PACS-detected LBGs at $z \sim 3$ in GOODS-North. We then obtain their total IR luminosities by integrating each best-fitted template in the rest-frame interval [8-1000] $\mu$m. The dust attenuation of the PACS-detected LBGs is parametrized by the ratio between the total IR and rest-frame UV luminosities \citep[see foe example][]{Buat2005}. Their total SFR, $SFR_{\rm total} = SFR_{\rm UV} + SFR_{\rm IR}$ , are calculated by combining the total IR and rest-frame UV luminosities and applying the \cite{Kennicutt1998} calibrations: $\textrm{SFR}_{\rm UV,uncorrected}[M_{\odot}\textrm{yr}^{-1}] = 1.4 \times 10^{-28}L_{1500}$ and $\textrm{SFR}_{\rm IR}[M_{\odot}\textrm{yr}^{-1}] = 4.5 \times 10^{-44}L_{\rm IR}$. It should be remarked that this procedure for obtaining the total SFR assumes that all the light absorbed by dust in the rest-frame UV is reemitted in turn in the FIR \citep[see for example][]{Magdis2010}. 

For comparison, and with the aim of carrying out evolutionary studies, we compare the FIR-derived properties of our PACS-detected LBGs at $z \sim 3$ with a sample PACS-detected LBGs at $z \sim 1$ in the GOODS-North and GOODS-South fields. At that redshift, the Lyman break is located in the UV and LBGs must be selected by employing UV colors from space-based observations. For segregating our LBGs at $z \sim 1$ we adopt the selection criterion of \cite{Oteo2013a}. The photometric redshifts, stellar mass, age, and dust attenuation of the galaxies are obtained from a SED-fitting procedure with BC03 to their UV to near-IR fluxes \citep{Capak2007} in the same way that it was done for LBGs at $z \sim 3$. Their FIR emission is also characterized with PACS data taken from the GOODS-\emph{Herschel} project. The FIR-derived properties of the PACS-detected LBGs at $z \sim 1$, i.e. dust attenuation and UV+IR-derived total SFR are obtained in the same way than it was done for the PACS-detected LBGs at $z \sim 3$. Due to the observational bias, LBGs at higher redshifts tend to be brighter. Therefore, if we want to compare galaxies at different redshifts which are selected through the same selection criterion, we must limit both samples to the same rest-frame UV luminosity. At $z \sim 3$, most LBGs have $\log{L_{\rm UV} / L_\odot} > 10$. Imposing this limit, we end up with a sample of 31 PACS-detected LBGs at $z \sim 1$.

\begin{sidewaystable}
\caption{\label{properties}Properties of the PACS-detected LBGs at $z \sim 3$ studied in the present work. We show the galaxy ID, sky position, photometric redshift, total IR luminosity, dust attenuation derived from the $L_{\rm IR}/L_{\rm UV}$ ratio \citep{Buat2005}, UV+IR-derived SFR \citep{Kennicutt1998}, stellar mass, and UV continuum slope.}
\centering
\begin{tabular}{lcccccccc}
\hline\hline
ID & RA [deg] & Dec [deg] & $z_{\rm phot}$ & $\log{\left( L_{\rm IR}/L_\odot \right)}$ & $A_{1200} \, [{\rm mag}]$ & $SFR_{\rm total} \, [M_\odot \, {\rm yr}^{-1}]$ & $\log{\left( M_*/M_\odot \right)}$ & UV slope \\
\hline
GOODS-MUSIC 15985 &    53.15745&    -27.70906&      2.71&     12.7&      5.3&    894.6&      8.5&     -1.50 \\
 GOODS-MUSIC 2989 &   53.21566&    -27.88574&      2.63&     12.4&      6.7&    436.4&     10.6&     -0.52\\
GOODS-MUSIC 4742 &    53.22782&    -27.86125&      2.69&     12.7&      5.9&    970.4&     11.6&     -0.74\\
GOODS-MUSIC 7180  &  53.18829&    -27.82934&      2.63&     12.4&      3.8&    518.0&     10.5&     -1.44\\
 GOODS-MUSIC 12229 &   53.09410&    -27.76094&      3.05&     12.7&      7.7&    987.3&     10.4&      0.17\\
GOODS-MUSIC 14620 &    53.12506&    -27.72682&      3.21&     12.5&      5.8&    594.9&     10.4&     -0.27\\
GOODS-MUSIC 17602 &    53.16990&    -27.68388&      2.77&     12.5&      5.3&    567.2&     11.4&     -1.14\\
GOODS-MUSIC 17746 &    53.11330&    -27.68094&      3.35&     13.0&      6.3&   1732.1&     10.8&    -0.08\\
GOODS-MUSIC 70202 &    53.11884&    -27.78282&      2.63&     13.1&      8.3&   2022.2&     11.3&      0.121\\
HDFN-1113 &   189.42146&     62.20577&      3.11&     12.9&      5.4&   1197.1&     10.5&     -0.66\\
HDFN-1871 &   189.24800&     62.37027&      2.63&     12.7&      4.4&    884.1&     10.6&     -0.95\\
HDFN-2128 &   189.43324&     62.30835&      2.54&     12.6&      5.8&    689.9&     11.3&     -0.99\\
HDFN-923 &   189.39473&     62.25430&      4.00&     12.7&      5.4&    970.2&     11.7&     -1.06\\
HDFN-1108 &   189.42360&     62.20658&      3.44&     13.0&      5.6&   1910.9&     11.6&     -1.39\\
HDFN-1520 &   189.18370&     62.21974&      3.11&     12.5&      3.6&    624.9&     10.4&     -1.32\\
HDFN-1700 &   188.91438&     62.21230&      3.11&     12.6&      5.4&    720.5&     10.6&     -0.33\\
\hline
\end{tabular}
\end{sidewaystable}

\begin{figure*}
\centering
\includegraphics[width=0.33\textwidth]{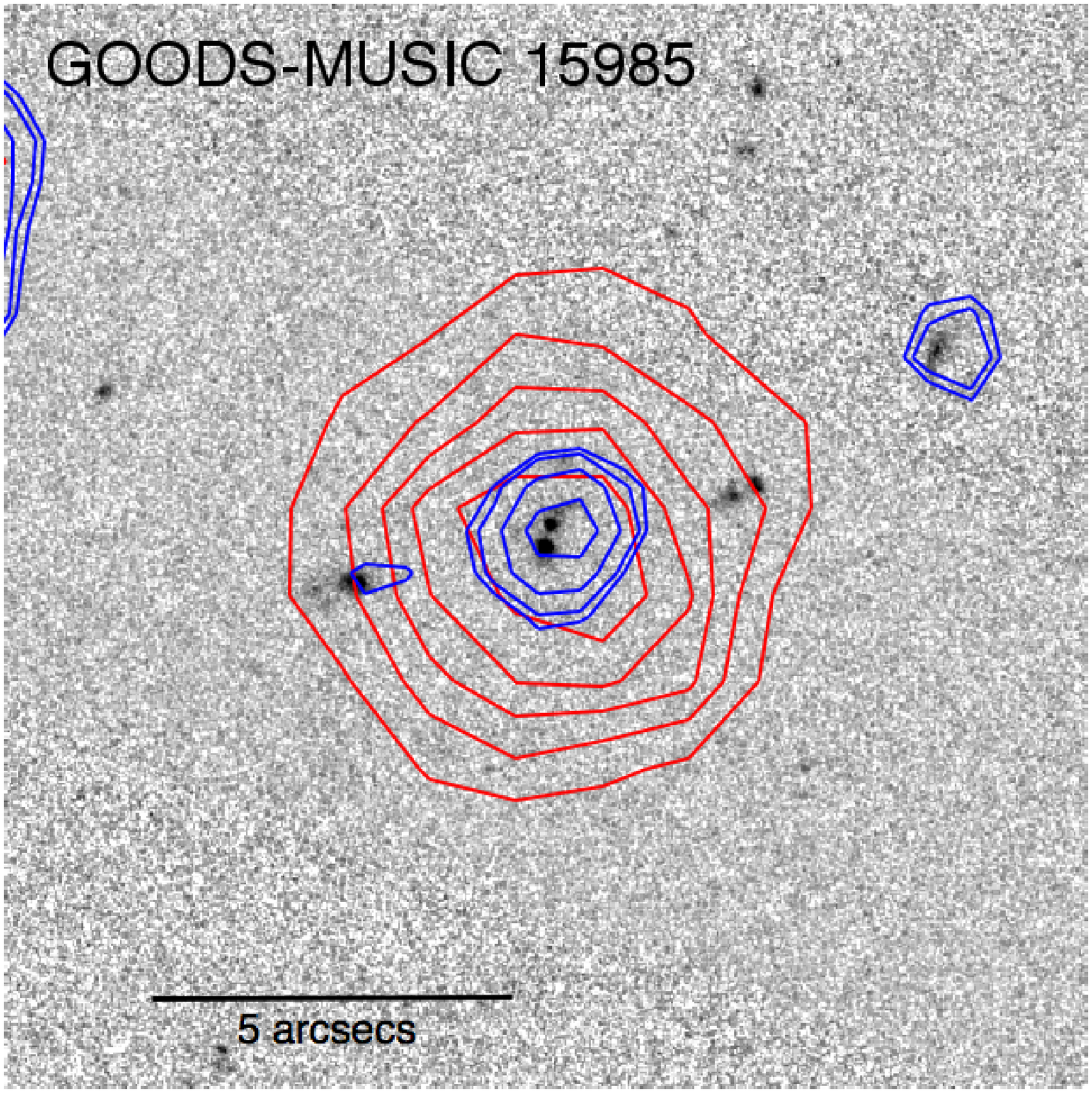}
\includegraphics[width=0.33\textwidth]{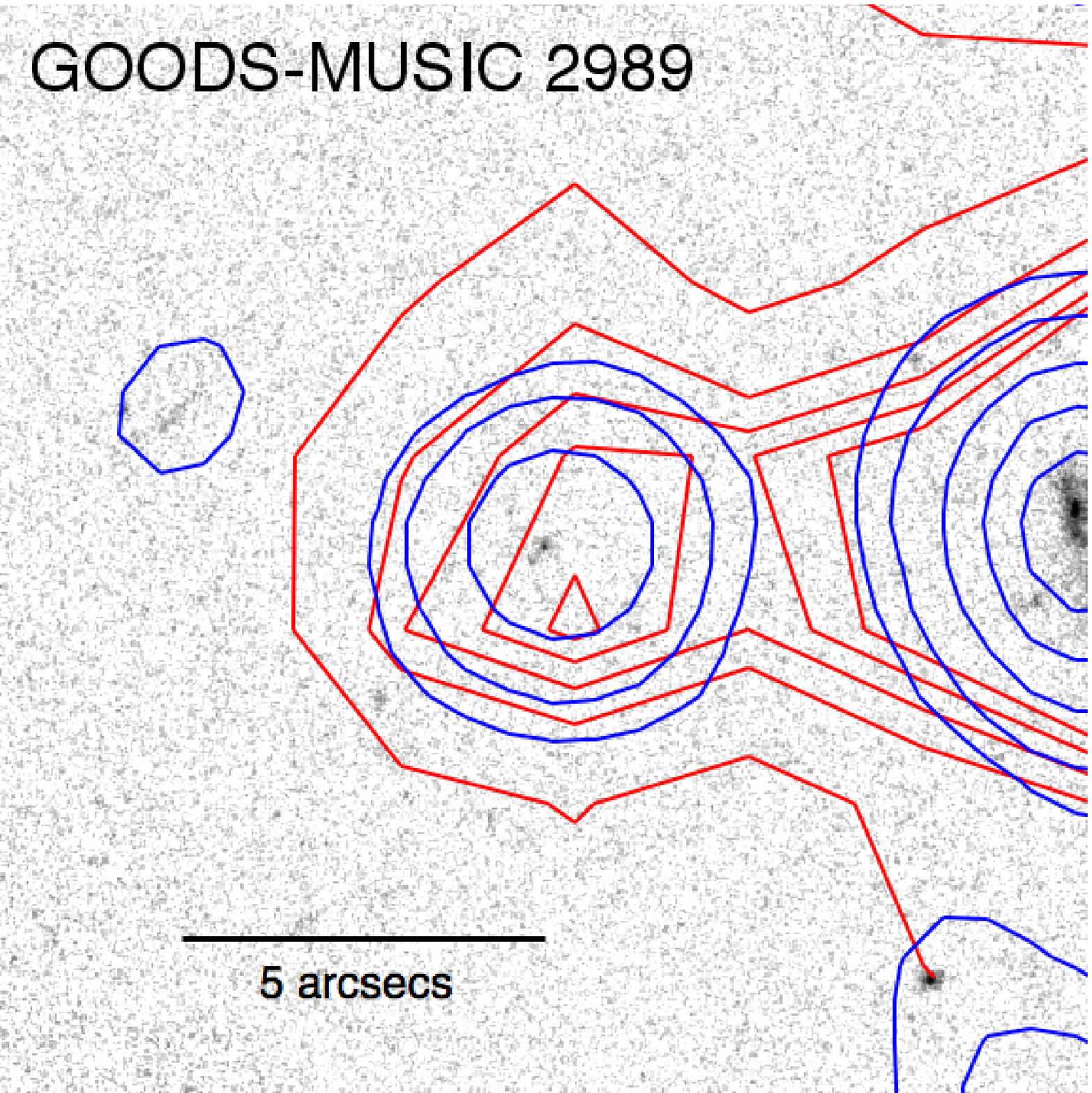}
\includegraphics[width=0.33\textwidth]{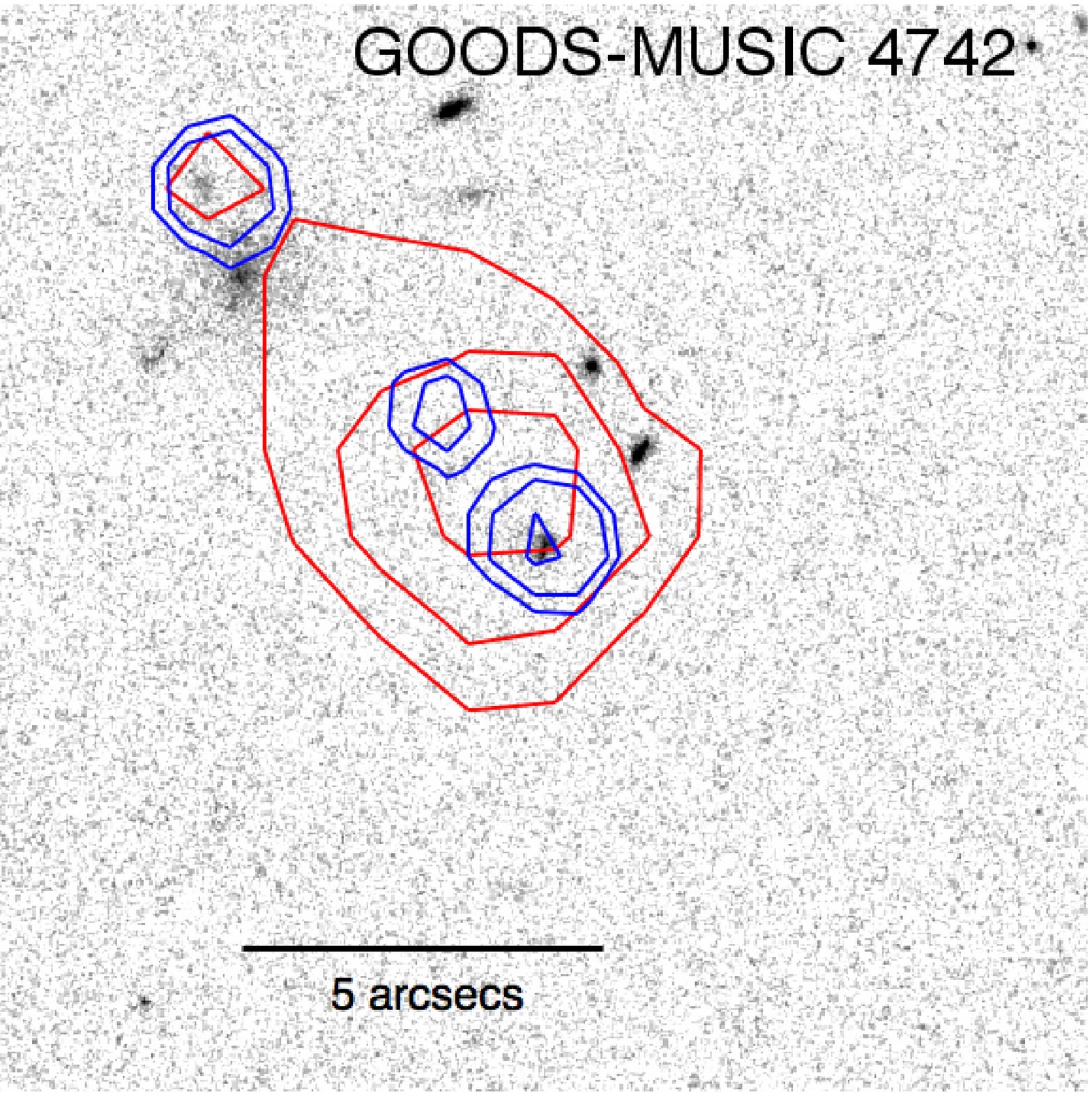}
\includegraphics[width=0.33\textwidth]{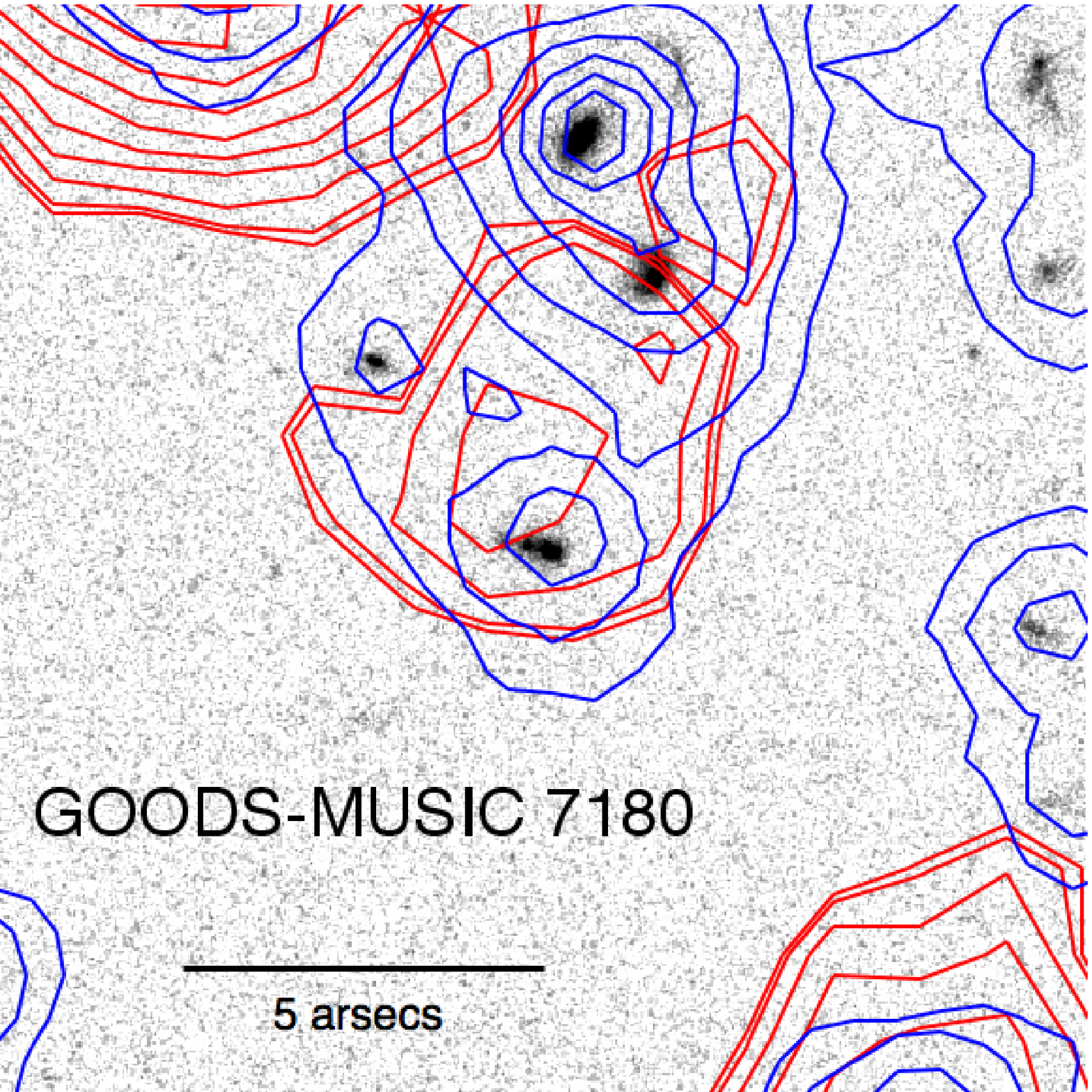}
\includegraphics[width=0.33\textwidth]{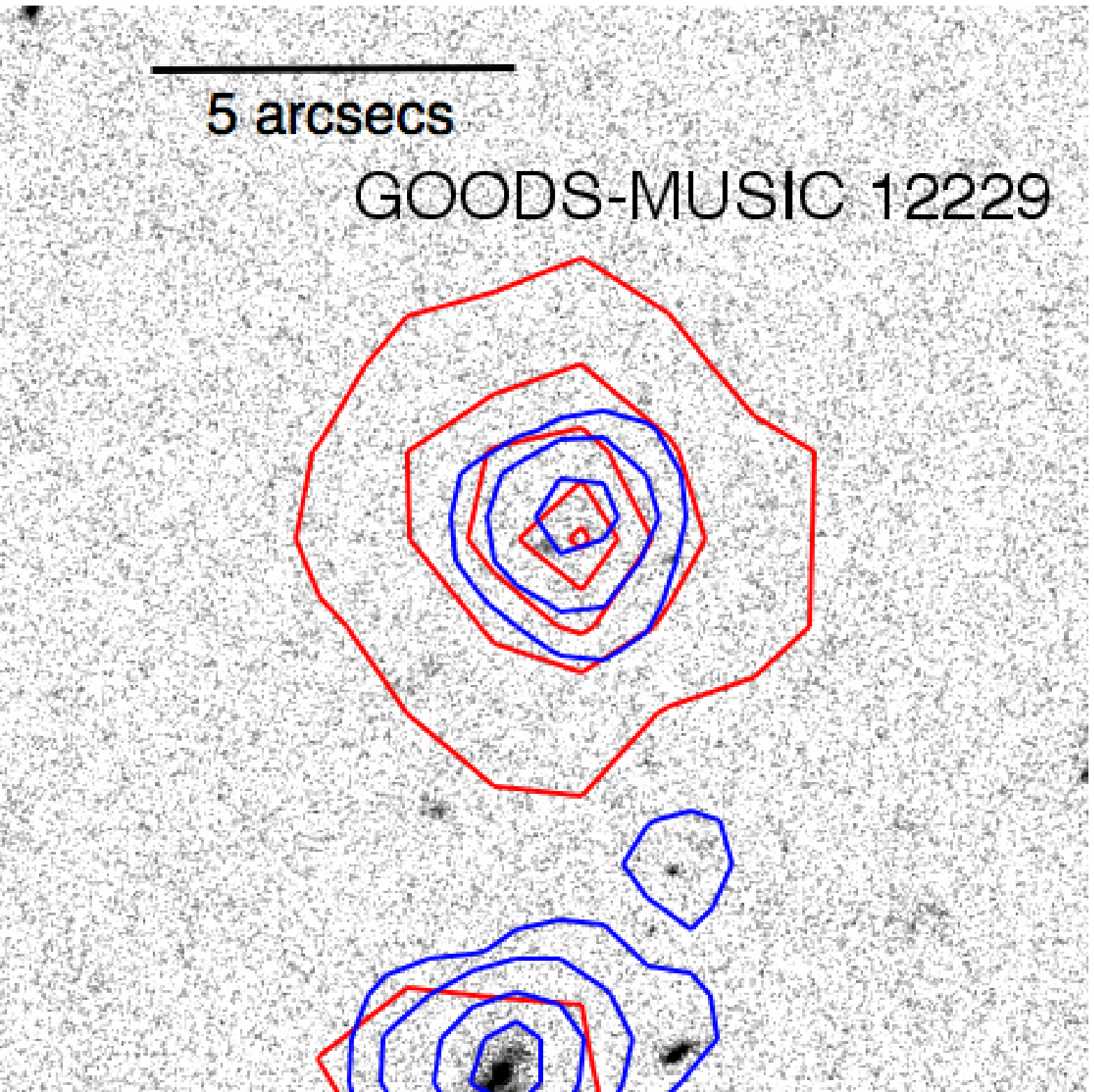}
\includegraphics[width=0.33\textwidth]{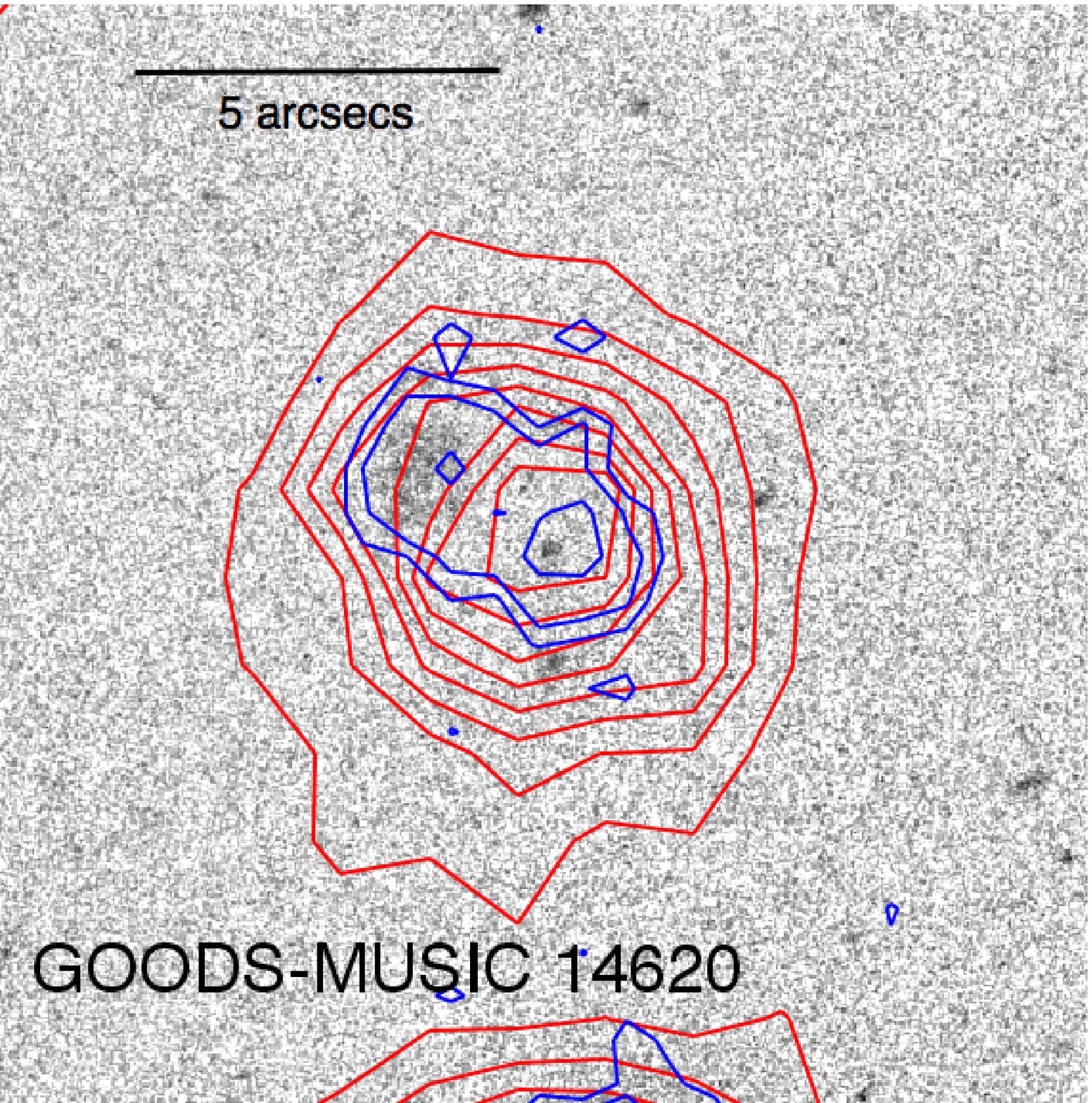}
\includegraphics[width=0.33\textwidth]{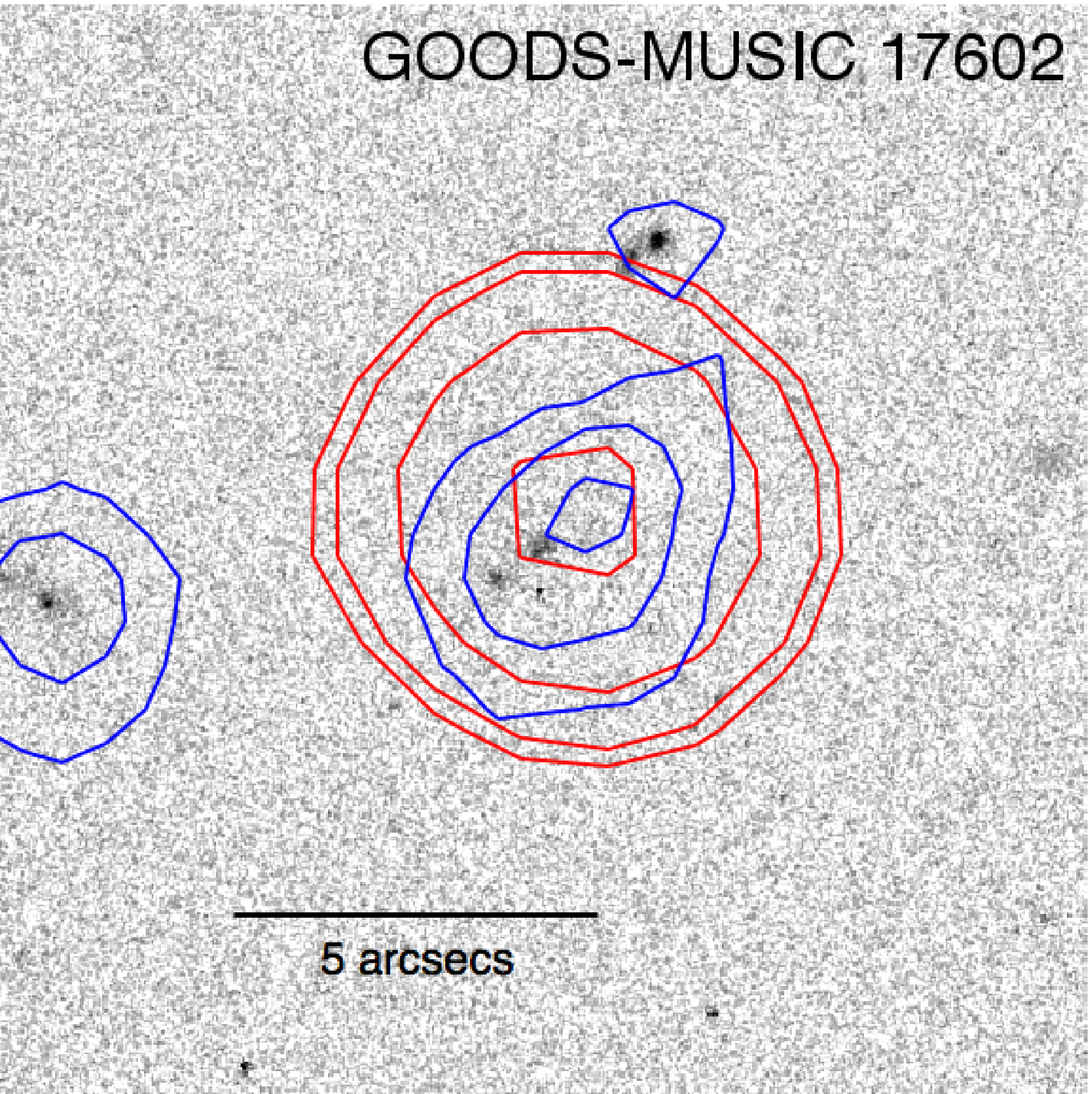}
\includegraphics[width=0.33\textwidth]{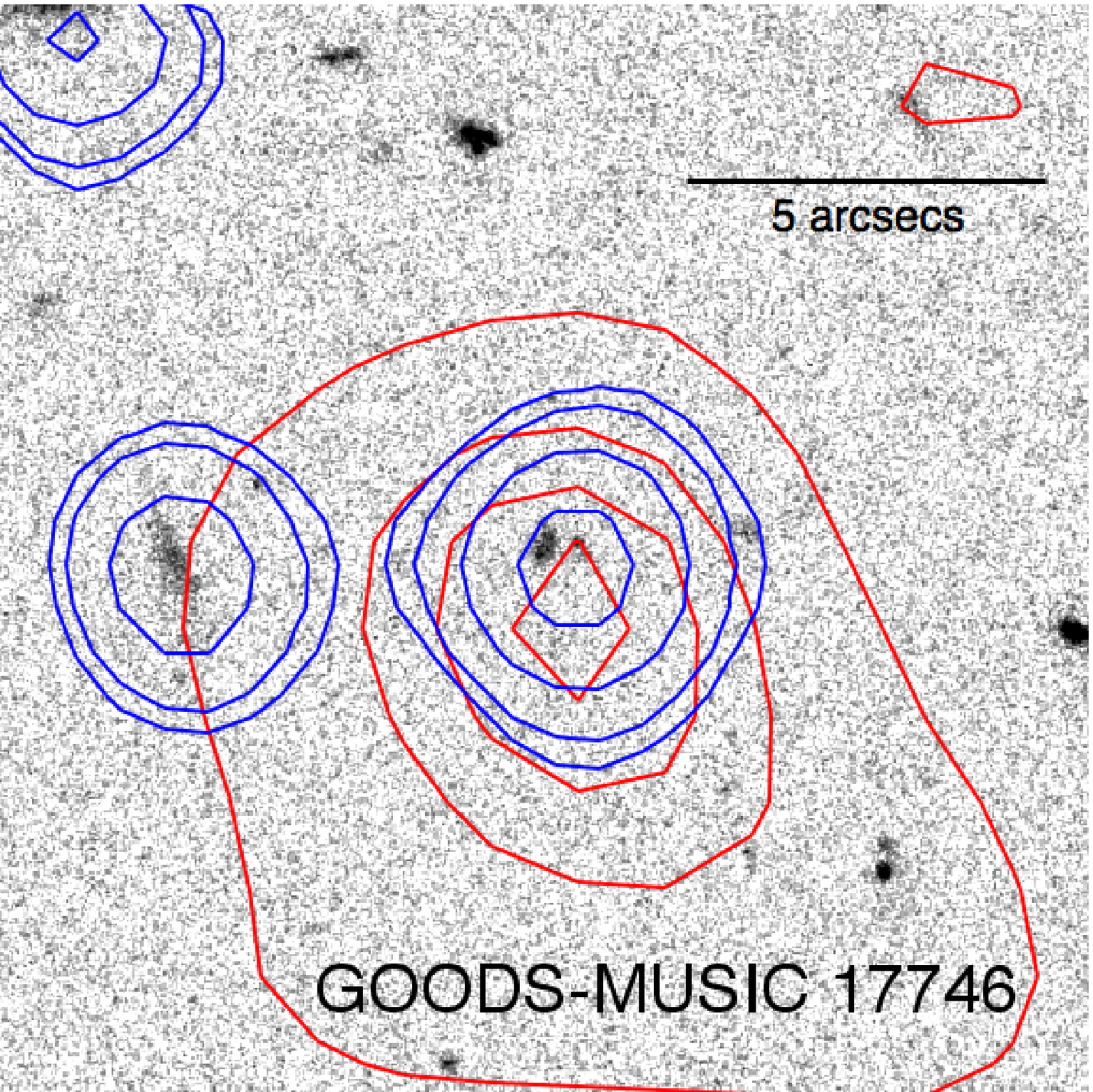}
\includegraphics[width=0.33\textwidth]{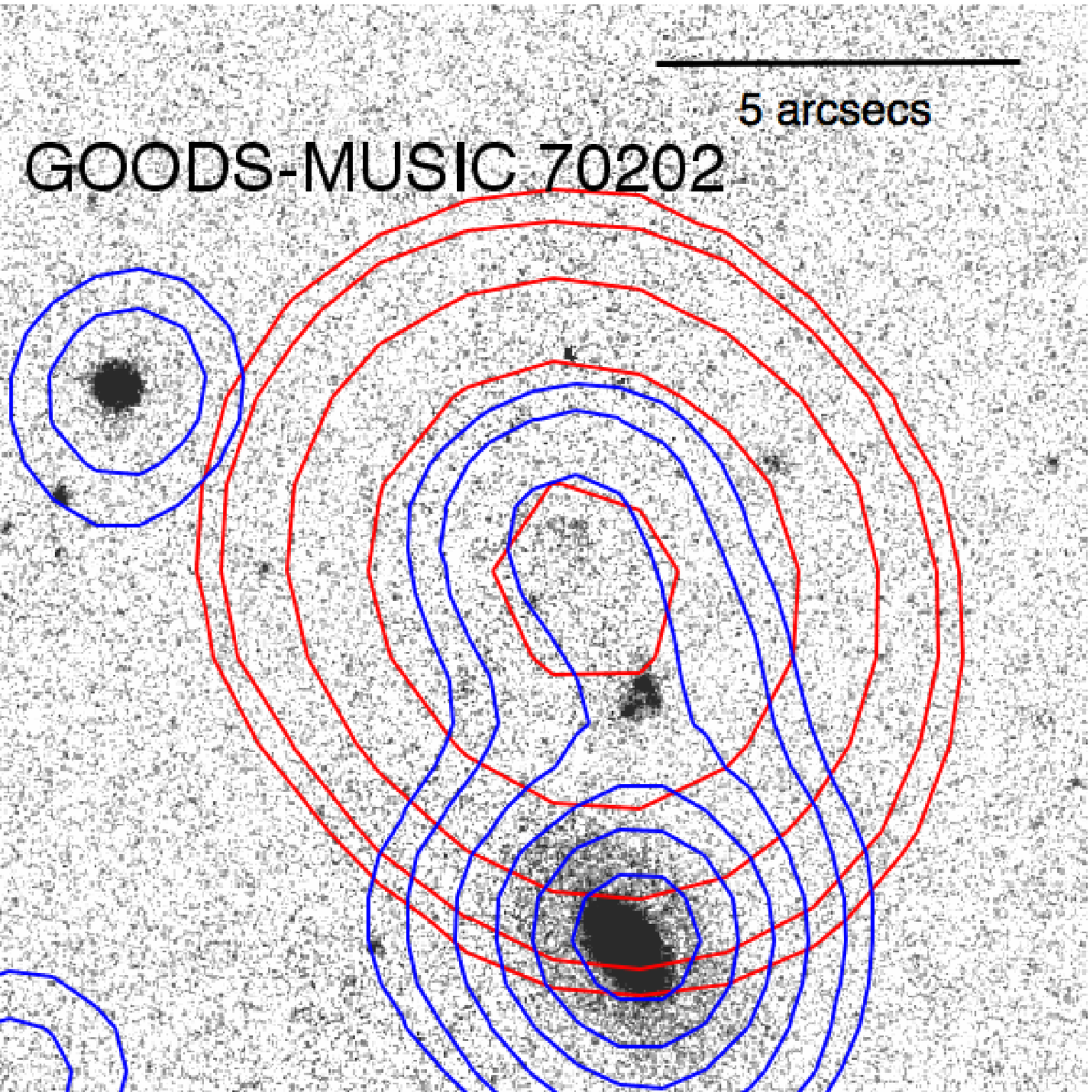}
\caption{ACS optical cutouts (15''x15'') of the nine PACS-detected LBGs located in the GOODS-South field. In each image, the PACS-detected LBG is the source located in the center. Blue contours represent the emission in the bluest available IRAC image. Red contours trace the MIPS-24$\mu$m emission. We use MIPS-24$\mu$m instead of PACS countours because they are visually clearer and the PACS fluxes are extracted with MIPS-24$\mu$m priors.
              }
\label{PACS_detected}
\end{figure*}

\end{appendix}

\end{document}